\documentclass[aps,prappl,reprint,groupedaddress]{revtex4-2}

\usepackage{amsmath,amssymb,graphicx}
\usepackage{epstopdf}
\usepackage{color}
\usepackage{MnSymbol}
\usepackage[version=3]{mhchem} 

\begin{document}


\title{Sub-diffraction-resolved spatial distribution of emitting excitons in STM-induced luminescence of 2D semiconductors via Richardson-Lucy deconvolution}



\author{Elys\'{e} Laurent} 
\affiliation{Universit\'{e} Paris-Saclay, CNRS, Institut des Sciences Mol\'{e}culaires d'Orsay, 91405, Orsay, France}
\author{Ricardo Javier Pe\~{n}a Rom\'{a}n}
\affiliation{Institute of Physics ``Gleb Wataghin'', Department of Applied Physics, State University of Campinas-UNICAMP, 13083-859, Campinas, Brazil}
\altaffiliation{Max Planck Institute for Solid State Research, Stuttgart 70569, Germany}
\altaffiliation{Technical University of Dresden, Institute of Solid State and Materials Physics,  Dresden 01069, Germany}
\author{Sarah Miller}
\affiliation{Universit\'{e} Paris-Saclay, CNRS, Institut des Sciences Mol\'{e}culaires d'Orsay, 91405, Orsay, France}
\author{Aditi Raman Moghe}
\affiliation{Institut de Physique et de Chimie des Mat\'{e}riaux de Strasbourg, Universit\'{e} de Strasbourg, CNRS, IPCMS, UMR 7504, F-67000 Strasbourg, France}
\author{Etienne Lorchat}
\affiliation{NTT Research, Inc., Physics \& Informatics (PHI) Laboratories, Sunnyvale, CA 94085, USA}
\author{S\'{e}verine Le Moal}
\author{Elizabeth Boer-Duchemin} 
\affiliation{Universit\'{e} Paris-Saclay, CNRS, Institut des Sciences Mol\'{e}culaires d'Orsay, 91405, Orsay, France}
\author{Luiz Fernando Zagonel}
\affiliation{Institute of Physics ``Gleb Wataghin'', Department of Applied Physics, State University of Campinas-UNICAMP, 13083-859, Campinas, Brazil}
\author{St\'{e}phane Berciaud}
\affiliation{Institut de Physique et de Chimie des Mat\'{e}riaux de Strasbourg, Universit\'{e} de Strasbourg, CNRS, IPCMS, UMR 7504, F-67000 Strasbourg, France}
\author{Eric Le Moal}
\affiliation{Universit\'{e} Paris-Saclay, CNRS, Institut des Sciences Mol\'{e}culaires d'Orsay, 91405, Orsay, France}
\email{eric.le-moal@universite-paris-saclay.fr}


\date{November 28, 2025}

\begin{abstract}
Using scanning tunneling microscopy-induced luminescence (STML), the optical properties of two-dimensional (2D) semiconductors may be investigated at the nanoscale. This is possible because the tunneling current under the tip is an extremely localized electrical excitation source. However, in most STML applications, the spatial distribution of the emission relative to the excitation point is unresolved. Yet this distribution contains key information about how the interaction of excitons with injected charge carriers affects the luminescence of these materials, and about exciton transport. Resolving this spatial distribution at the nanoscale is relevant both for a fundamental understanding of exciton physics and for device applications; yet it remains a significant challenge. In this work, we resolve the spatial distribution of the emission beyond the diffraction limit of light by deconvolving real-space optical microscopy images of the STML using an iterative algorithm, i.e., Richardson-Lucy (RL) deconvolution. To showcase this technique, we apply it to the STML of monolayer tungsten diselenide (\ce{WSe_2}) and tungsten disulfide (\ce{WS_2}). Thus, we highlight hitherto ignored or misunderstood aspects of STML on 2D semiconductors related to exciton and charge carrier transport, namely the dependence of the spatial distribution of emission on the tunnel current setpoint and the origin of the emission from hot spots located micrometers from the excitation source.

\end{abstract}


\maketitle


\section{Introduction}

The two-dimensional (2D) geometry of monolayer van der Waals materials is the source of their unique optical and electronic properties~\cite{Mak2010,Splendiani2010,Geim2013,Novoselov2016,Brar2017,Binder2017}. These properties are extremely sensitive to their immediate environment and can be tuned to an unprecedented extent via external stresses and fields~\cite{Branny2017,Xie2018,Qi2023}. Such control opens up exceptional possibilities for applications in the fields of ultrasensitive sensors~\cite{Anichini2018,Meng2019,Pang2020,Khan2021,Rohaizad2021} and tunable optoelectronic devices~\cite{Wang2012,Cheng2014,Furchi2014,Ross2014,Akinwande2014,Clark2016,Jariwala2016,Mak2016,Liu2017,Brar2018,Lemme2022}.Having materials of near-atomic thickness also offers new opportunities for fundamental research. In particular, the effects of quantum confinement and reduced dielectric screening on the dynamics of excitons (Coulomb-bound electron-hole pairs) have been extensively studied in 2D semiconductors such as monolayer transition metal dichalcogenides (TMDs)~\cite{Palacios-Berraquero2017,Raja2017,Wang2018,Unuchek2018,Jauregui2019,Baek2020,PereaCausin2022,Tagarelli2023,Malic2023}. Furthermore, the 2D geometry of these materials makes it possible to directly probe and manipulate elementary optical and electronic processes in real space at the nanoscale. This may be carried out using a near-field technique such as scanning probe microscopy (SPM)~\cite{Koo2024}. Atomic force microscopy combined with photoluminescence (PL) spectroscopy has been used to locally control exciton excitation and emission processes in monolayer TMDs on the nanoscale via tip-induced electromagnetic, electronic and mechanical stress effects~\cite{Park2016,Park2018,He2019,Darlington2020,Zhang2022,Kim2024}. Scanning tunneling microscopy (STM) is also increasingly used in combination with optical spectroscopy to locally induce electroluminescence from monolayer TMDs via inelastic tunneling effects and to correlate emission properties with nanoscale topography~\cite{Krane2016,Pommier2019,Schuler2020,Pechou2020,PenaRoman2020,PenaRoman2022a,Ma2022,ParraLopez2023,Geng2024,Huberich2025}.

However, what these techniques combining SPM and optical detection produce are in the vast majority of cases optical signal maps. To obtain such a map, the tip scans the sample while the emitted light, spatially integrated over the entire field of view, is recorded. The resulting maps are fundamentally different from optical microscopy images. Although the tip-sample interaction is localized on the nanoscale, the obtained maps provide no information on the spatial distribution of emitters. This drawback concerns most of the SPM-based optical experiments on TMDs reported so far. The exceptions are those based on the combination of SPM with wide-field optical microscopy, where an optical microscopy image is recorded~\cite{Pommier2019,PenaRoman2022a,PenaRoman2022}. However, without dedicated image analysis, resolving the spatial distribution of emitters using diffraction-limited imaging tools is challenging, especially when exciton diffusion lengths are shorter than the resolution limit of the optical microscope. Iterative deconvolution algorithms based on empirical knowledge of the point spread function (PSF) have been widely used, e.g., to improve the spatial resolution of conventional fluorescence microscopy images beyond the diffraction limit in life sciences~\cite{Kempen1997,Verveer1999,Sibarita2005,Sage2017}. However, these algorithms have never been applied to experiments coupling SPM and optical microscopy so far, presumably due to the difficulty of defining and measuring the PSF in such SPM-based experiments.

In the present article, we investigate the STM-induced excitonic luminescence (STML) of monolayer tungsten diselenide (\ce{WSe_2}) and tungsten disulfide (\ce{WS_2}) in an air-operated STM coupled to an optical microscope, and we analyze the resulting real-space optical microscopy images in order to resolve the spatial distribution of radiatively recombining excitons around the excitation source. The image analysis involves deconvolving the optical microscopy images by the theoretical PSF of the optical microscope using an iterative algorithm, i.e., Richardson-Lucy (RL) deconvolution~\cite{Richardson1972,Lucy1974}. In the present case, the PSF is well defined because the emitters are identified as spin-allowed bright (neutral and charged) excitons in the TMD monolayer, thanks to their emission spectrum~\cite{Pommier2019,PenaRoman2022a} (see the STML spectra shown in Figure S1 of the Supporting Information). Thus, the PSF is numerically calculated by simulating the image of an in-plane electric dipole source emitting circularly polarized photons and by taking into account the experimentally measured spectral distribution of the emitted light. We use this method to demonstrate the current-dependence of the spatial distribution of emitters around the excitation source in monolayer \ce{WSe_2} and to resolve the emission from hot spots microns away from the excitation source in monolayer \ce{WS_2}.

\section{Methods}

\subsection{STM and optical experiments} 

A commercial STM head from JPK Instruments (NanoWizard 3) is mounted on an inverted optical microscope (Nikon Eclipse Ti-U) which is equipped with an oil-immersion objective (Nikon CFI Apochromat $100\times$ $1.49\mathrm{NA}$ TIRF objective)~\cite{Cao2017}. The resulting light from STML and PL experiments is collected through the transparent substrate. Real-space images are recorded using a water-cooled charge-coupled device (CCD) camera (Andor iKon-M). PL from the sample is excited using a continuous-wave argon-ion laser emitting at a wavelength of $465.8$~nm, under wide-field episcopic illumination in normal incidence. The laser power at the sample is about $0.01-0.02$~W~cm$^{-2}$. A longpass filter from $\lambda=491$~nm is used. The STM tips used are prepared by electrochemical etching of tungsten wires and have a radius of curvature at the apex of approximately $30$~nm (see Fig.~S2 in Ref.~\cite{LeMoal2013}).

\subsection{PSF calculation}

Spin-bright excitons in semiconducting monolayer TMDs have their transition dipole moment oriented in the plane of the monolayer and their radiative recombination yields circularly polarized photons~\cite{Xiao2012,Cao2012}. To model the emission of circularly polarized photons, we coherently sum the emission of two in-plane electric dipoles oriented perpendicular to each other and phase-shifted by $\pi/2$, which are located at the same point and oscillate at the same frequency. The oscillation frequency in the model corresponds to the radiative emission of neutral excitons in the experiment. The substrate is modeled as an $85$~nm thick dielectric layer, corresponding to the ITO layer, on a semi-infinite space that has the dielectric permittivity of glass (i.e., $2.28$). For the dielectric layer corresponding to ITO, the dielectric permittivity used is $2.48+0.066i$ (measured at $\lambda_0=750$~nm by ellipsometry on our ITO-coated glass slides) when simulating the experiments conducted on \ce{WSe_2}, and $2.95+0.054i$ (measured in the same way at $\lambda_0=610$~nm) when dealing with \ce{WS_2}. In the model, the point source is located at a distance of $1$~nm from the air-substrate interface. The spatial-frequency distribution of the complex electric field vector in the far field is calculated using Fresnel coefficients and a plane-wave decomposition of the dipole emission (see Chap.~10 in Ref.~\cite{Novotny2006}). A spatial-frequency short-pass filter is applied to this distribution in order to simulate the limited angular acceptance of the microscope objective (NA~$= 1.49$). This spatially-filtered distribution is Fourier-transformed and we calculate its square modulus to obtain the simulation of the real-space optical microscopy image of a single emitter, i.e., the PSF of the microscope for the specific case of this experiment. To account for the spectral width of the emission, we consider a weighted sum of real-space images calculated at different oscillation frequencies, where the weighting factors of the sum and the oscillation frequencies are obtained from the experimentally measured STML spectra, shown in Supporting Information for \ce{WSe_2} and taken from a previous work~\cite{PenaRoman2022a} for \ce{WS_2}. 

\subsection{Image deconvolution}

In general, experimental optical images are both blurred, due to convolution of the object with the PSF of the imaging system, and noisy, due to the photon statistics of the light source and electronic noise of the detection device. Given a blurred and noisy image $I(x,y)$, RL deconvolution algorithms estimate the most likely object $O(x,y)$, knowing the PSF (noted $H$ below) and the statistical noise distribution~\cite{Richardson1972,Lucy1974}. In the absence of noise, the image may be expressed as $I=O\ast H$, where $\ast$ stands for the 2D convolution. Assuming that the noise has Poisson distribution and is spatially uncorrelated, the statistics of the noisy image, i.e., the probability that it equals $I$ knowing $O\ast H$, reads:
\begin{equation}
	P(I|O\ast H)=\prod_{x,y} \frac{(O\ast H)^I}{I!} \cdot e^{-O\ast H}
\end{equation}
where $\prod_{x,y}$ is the product over all pixels of the image and $I!$ is the factorial of $I$, i.e., of the number of counts measured in each pixel of the image $I(x,y)$. Maximizing the likelihood of the object estimate comes to maximize $P(I)$, or equivalently, to minimize a functional $J(O)=-\ln \left(P(I|O\ast H)\right)$, which reads:
\begin{equation}
	J(O)=\sum_{x,y} O\ast H - I \cdot \ln (O\ast H)
\end{equation}
where $\sum_{x,y}$ is the sum over all pixels of the image. $J$ being a convex function of $O$, its minimum corresponds to a zero of its derivative $\nabla J$, which reads:
\begin{equation}
	\nabla J(O)=\left[ 1-\frac{I}{O\ast H}\right]\ast H^T 
\end{equation}
where $H^T(x,y)=H(-x,-y)$. We use the so-called multiplicative solution of the RL deconvolution problem, which uses the fact that the estimated object is expected not to change between iteration $n$ and $n+1$ once the algorithm has converged, i.e., $O_n = O_{n+1}$. This yields the following iterative solution:
\begin{equation}\label{eq1}
	O_{n+1}=O_n \cdot\left[ \left( \frac{I}{O_n\ast H}\right)  \ast H^T \right]
\end{equation}
Alternatively, Eq.~\ref{eq1} can be written as follows, where the use of the Fourier transform replaces the 2D convolution with products~\cite{Liu2025}:
\begin{equation}\label{eq2}
	O_{n+1}=O_n \cdot\mathcal{F}^{-1}\left[ \mathcal{F}\left( \frac{I}{\mathcal{F}^{-1}\left[ \mathcal{F}(O_n)\cdot \mathcal{F}(H)\right] }\right)\cdot\mathcal{F}\left(H^T\right)  \right]
\end{equation}
where $\mathcal{F}$ and $\mathcal{F}^{-1}$ stand for Fourier transform and inverse Fourier transform, respectively. In this work, we use two different versions of the RL algorithm, based on Eq.~\ref{eq1} (script~1) and Eq.~\ref{eq2} (script~2), respectively. Both scripts are programmed in Python and are adapted from code available on the GitHub platform~\cite{deconvolution,Stallinga2024}. We show that using one version or the other does not yield the same computational artifacts and that using script~2 yields better results than script~1 for the deconvolution of upsampled images. The Python code used, as well as all the raw data corresponding to the results presented in this article, are freely available online~\cite{LeMoal2025a}.

\section{Results and discussion}

\subsection{Demonstration of principle}

%
\begin{figure*}
	\includegraphics[width=1.0\linewidth]{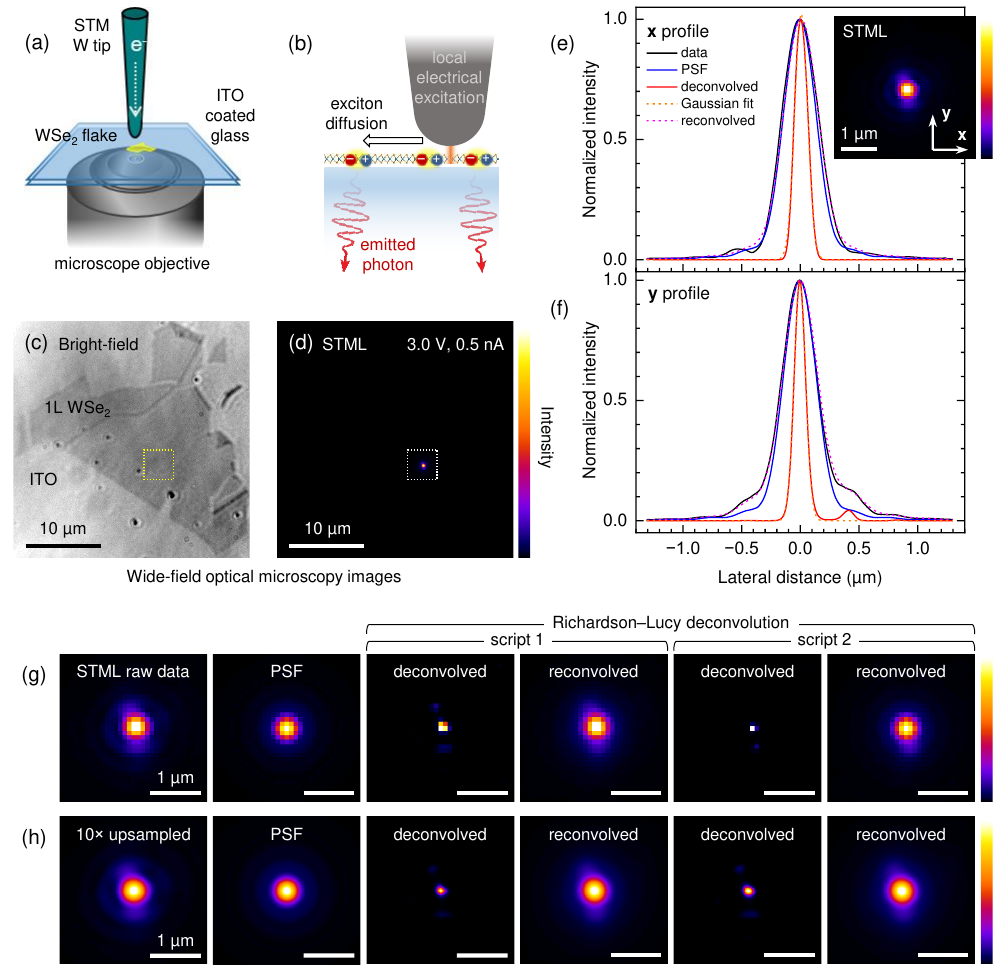} 
	\caption{(a) Schematic of the experiment: the sample is a \ce{WSe_2} flake on an ITO-coated glass coverslip, placed between the tungsten tip of an STM and a high-NA microscope objective. (b) Schematic of the local electrical excitation of excitons in monolayer \ce{WSe_2} using the tunneling current from the STM tip, and the diffusion of these excitons before they radiatively decay. (c) White-light transmission optical microscopy of the sample: a flake of monolayer (1L) \ce{WSe_2}. (d) Optical microscopy image of the STM-induced luminescence (STML) of the same area as in (c): the STM tip is located at a fixed lateral position, i.e., where the bright spot is observed, and is in the tunneling regime (sample bias $V_\mathrm{s}=3.0$~V, current setpoint $I_\mathrm{t}=0.5$~nA). (e) Intensity profiles obtained from the image shown in (d) (black curve), a numerical simulation of the point spread function (PSF) (blue curve), the image deconvolved using the Richardson-Lucy algorithm (red curve) and fitted using a Gaussian function (dashed orange curve), and the 2D convolution of the deconvolved image with the PSF (dashed magenta curve) along the $\mathbf{x}$ axis. Inset: zoomed image of the area delineated by a dotted line in (d). (f) Same as in (e) along the $\mathbf{y}$ axis. (g) Zoomed images of (from left to right) the STML data, the simulated PSF, the deconvolved image, and the 2D convolution of the latter with the PSF, using two different versions of the Richardson-Lucy algorithm (see Methods). In (g), the native pixel sampling of the STML image is used. (h) Same as (g), where a $10\times$ upsampled version of the STML image is used.}
	\label{FIG-1}
\end{figure*}

Figures~\ref{FIG-1}(a) and \ref{FIG-1}(b) schematically show the principle of the experiment. The experimental setup consists of an air-operated STM head mounted on top of an inverted optical microscope~\cite{Cao2017}. The sample is a mechanically exfoliated TMD microflake deposited onto an indium tin oxide (ITO)-coated glass coverslip using a dry transfer method~\cite{Castellanos-Gomez2014} (ITO thickness $85$~nm). A bright-field microscopy image of a monolayer area of a \ce{WSe_2} sample is shown in Figure~\ref{FIG-1}(c). The STM tip is an electrochemically etched tungsten wire. The emitted light is detected in transmission through the substrate using an oil-immersion microscope objective that has a high numerical aperture (NA~$= 1.49$). As represented in Figure~\ref{FIG-1}(b), the electrical excitation due to the tunneling current is extremely localized. However, the excitons created by the inelastic tunneling current may diffuse away from the excitation source before radiatively decaying~\cite{Mouri2014,Kato2016,Kulig2018}. Photon mapping, i.e., the most widely used imaging mode in STML experiments~\cite{Kuhnke2017}, is insensitive to this spatial distribution of emitters (i.e., of emitting excitons) around the excitation source. In photon mapping, where the tip scans the sample while the optical signal is simultaneously recorded, the total number of detected photons at each position of the tip is assigned to a pixel of the photon map, irrespective of the position of the emitters. This is why we used a different technique, i.e., wide-field optical microscopy, to visualize the spatial distribution of the emitted light in real-space~\cite{Pommier2019}. 

Figure~\ref{FIG-1}(d) shows the real-space optical microscopy image of the light emitted from the \ce{WSe_2} sample when it is excited by the tunneling current from the STM tip located in the center of the square monolayer area delineated by a dotted line. The STM tip is in the tunneling regime in constant-current mode, at a sample bias voltage of $3$~V and a setpoint current of $0.5$~V, and is not scanning. The inset in Figure~\ref{FIG-1}(e) shows a zoomed image of the STML emission spot seen in the optical microscopy image shown in Fig.~\ref{FIG-1}(d). Figures~\ref{FIG-1}(e) and~\ref{FIG-1}(f) show the spatial distribution (black curve) of the emitted light along a line crossing this emission spot along the $\mathbf{x}$ and $\mathbf{y}$ axes, respectively. The shape and width of the emission spot in the optical microscopy image results from the convolution of the spatial distribution of the emitters by the point spread function (PSF) of the optical microscope. Thus, the spatial distribution of the emitters may be retrieved from an image deconvolution procedure. 

In order to take into account the frequency and polarization-dependence of the PSF, we model the PSF of the optical microscope as the image of a single point source that has the emission properties of a spin-bright exciton in monolayer \ce{WSe_2} (see Methods). To deconvolve the experimental optical microscopy images with the PSF discussed above, we use a type of iterative algorithm, i.e., Richardson-Lucy (RL) deconvolution~\cite{Richardson1972,Lucy1974}, which was initially developped in astronomy and has since been widely used in the life sciences to enhance spatial resolution in fluorescence microscopy~\cite{Holmes1989,Dey2006,Laasmaa2011,Mukamel2012,Wang2014b,Stroehl2015,Perez2016,Zhang2019} (see Methods). 

Figure~\ref{FIG-1}(g) shows a zoomed version of the experimental image shown in Fig.~\ref{FIG-1}(d) and of the simulated PSF of the optical microscope, both plotted with the native pixel sampling of the experimental image. The size of a pixel is about $87$~nm, which is smaller than the resolution limit or PSF width of the optical microscope. Abbe diffraction limit is equal to $\lambda_0/2 \mathrm{NA} = 0.25~\mu$m at the considered emission wavelength ($\lambda_0=748$~nm, $\mathrm{NA}=1.49$). The results obtained after $280$ iterations of two different versions of the RL deconvolution algorithm are also shown. In the first script, for each iteration of the algorithm the 2D convolution of the PSF with an estimate of the object is computed. In the second script, the 2D convolution is replaced by the product of the Fourier transforms of the PSF and object estimate
(see Methods). The goal of this comparison is to identify which of the two scripts is less prone to artifacts and convergence issues. To evaluate the performance of the deconvolution, the deconvolved images are convolved once again with the PSF (i.e., the inverse operation of deconvolution is performed) and compared to the initial experimental image. This operation reveals that on images with native pixel sampling and for the iteration number considered, RL deconvolution using the first script yields a more accurate result than using the second script. This is because the latter method underestimates the width of the emitter distribution. Moreover, for both methods, we find that the native pixel sampling of the experimental image is insufficient to accurately determine the deconvolved images and estimate the width of the emitter distribution. This is especially true for RL deconvolution using the second script, where the deconvolved image shown in Fig.~\ref{FIG-1}(g) consists of a single bright pixel surrounded by a few near-dark pixels. In such an undersampled image, the accurate estimation of a distribution width via the fitting of intensity profiles with (e.g.) Gaussian functions is not possible. Below, we demonstrate that the estimation of the emitter distribution width may be improved by applying the RL deconvolution to $10\times$ upsampled versions of the experimental images and the PSF. The upsampled data is obtained by data interpolation using the ImageJ software~\cite{Schneider2012}. 

Figure~\ref{FIG-1}(h) shows a set of images similar to Figure 1(g), except that the experimental STML data and the simulated PSF are upsampled by a factor of $10$. Starting from upsampled images, the two scripts yield emitter distributions that are much more similar to each other as compared to images with native pixel sampling when considering the profile of these distributions along the $\mathbf{x}$ and $\mathbf{y}$ axes, as shown in Figures 1(e) and 1(f) (red curve). Moreover, due to the finer sampling, a more accurate estimation of the distribution width by fitting a Gaussian function is possible, as shown in Figures 1(e) and 1(f) (orange dashed curve). Along the $\mathbf{x}$ and $\mathbf{y}$ axes, the standard deviations of the emitter distributions estimated from the two versions of the RL algorithm differ by less than $10\%$. Moreover, for both scripts used, the RL deconvolution reveals that the asymmetric tail of the STML spot in the undeconvolved image results from the presence of two secondary emission spots $0.4$ to $0.5~\mu$m from the lateral tip position. However, we find a convergence issue for the RL algorithm using the first script, which yields that the central spot in the deconvolved image tends to take a cross shape beyond a certain number of iterations, as can be observed in Figure 1(h) and in the zoomed version of the image shown in Figure~S2 of the Supporting Information. This convergence issue and resulting artifact are not observed for the RL algorithm using the second script. Using the latter algorithm, from the Gaussian fit of the central spot we find that the estimated distribution of emitters has a full width at half maximum (FWHM) equal to about $0.12~\mu$m along the $\mathbf{x}$ axis and $0.13~\mu$m along the $\mathbf{y}$ axis. Note that these values of FWHM correspond to less than two pixels in the raw data (in the native pixel sampling) and about half the Abbe diffraction limit $\lambda_0/2 \mathrm{NA} = 0.25~\mu$m at the considered emission wavelength. Now that we have demonstrated our approach in principle, we use it below to highlight some poorly understood aspects of STML on 2D semiconductors, which are related to exciton and charge carrier diffusion.

\subsection{Applications}

\subsubsection{Spatial distribution of emitters versus tunnel current}

%
\begin{figure}
	\includegraphics[width=1.0\linewidth]{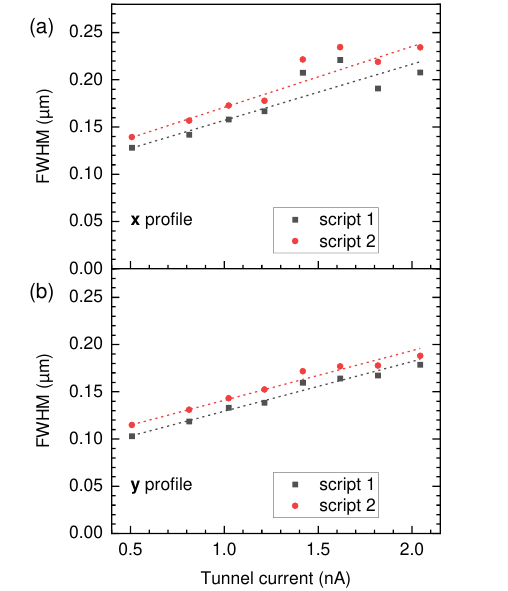} 
	\caption{Effect of the tunneling current on the spatial distribution of emitters in STML experiments on monolayer \ce{WSe_2}. The full width at half maximum (FWHM) of this distribution is determined by fitting Gaussian intensity profiles to deconvolved optical microscopy images of STML measured for various current setpoints on the same sample area. The obtained FWHM is plotted as a function of the current setpoint. Intensity profiles obtained from the same deconvolved images along the $\mathbf{x}$ and $\mathbf{y}$ axes are considered in (a) and (b), respectively. Two different versions of the Richardson-Lucy deconvolution algorithm are compared. In the first script (black squares) 2D convolution between object and PSF is computed at each iteration, while in the second script (red dots) this operation is replaced by the product of object and PSF Fourier transforms (see Methods). The dashed lines are linear fits of the data. Sample bias is $2.8$~V for all data.}
	\label{FIG-2}
\end{figure}

Figure~\ref{FIG-2} highlights the dependence of the spatial emitter distribution on the current setpoint in STML experiments on TMD monolayers. Optical microscopy images of STML are obtained using different current setpoints (but the same bias voltage $V_\mathrm{s}=2.8$~V) on the same area of the monolayer \ce{WSe_2} sample as shown in Figure~\ref{FIG-1}. These images are then deconvolved using the RL algorithms described above after $10\times$ upsampling. The FWHM of the estimated emitter distribution, obtained from the Gaussian fit of profiles obtained along the $\mathbf{x}$ and $\mathbf{y}$ axes, is plotted as a function of the measured tunneling current (which is equal to the setpoint current to within $2\%$). Figure~\ref{FIG-2} reveals that the width of the emitter distribution around the excitation source increases with the tunneling current amplitude.
At a setpoint current of $0.5$~nA, FWHMs similar to those measured in Figure~\ref{FIG-1} (i.e., $0.12~\mu$m and $0.13~\mu$m) for the same current are found, namely $0.13~\mu$m along the $\mathbf{x}$ axis and $0.10~\mu$m µm along the $\mathbf{y}$ axis using the first script, and $0.14~\mu$m and $0.12~\mu$m using the second script, respectively. At $2.0$~nA, an increase of $65$ to $80\%$ of these FWHMs is observed. The analysis of the data by linear fit shows that to a first approximation the FWHM increases linearly with the tunneling current with a slope between $0.05$ and $0.06~\mu$m nA$^{-1}$ (depending on the axis considered). 

Based on order of magnitude reasoning, we hypothesize the following about the origin of the dependence on the current observed in Fig.~\ref{FIG-2}. For the STM parameters used, the quantum efficiency of STML on this sample is of the order of $10^{-6}$ photons per tunneling electron. The radiative quantum yield of monolayer semiconductor TMDs on ITO in ambient air, previously measured in PL~\cite{PenaRoman2022}, is of the order of $10^{-3}$ to $10^{-2}$ (emitted photons per absorbed photon). We assume that an exciton is created for each absorbed photon in the PL measurements. Therefore, in the STML experiments, we estimate that $10^{-4}$ to $10^{-3}$ excitons (neutral or charged) are generated per tunneling electron under the STM tip, which at the considered currents corresponds to $10^{6}$ to $10^{7}$ excitons per second. Given the sub-nanosecond lifetime of bright excitons, we can exclude that the broadening of the emitter distribution with current under the STM tip is due to exciton-exciton interactions, in contrast to focused laser PL experiments\cite{Kulig2018} (where the density of generated excitons is several orders of magnitude higher). Instead, we propose that the effect observed in Fig.~\ref{FIG-2} results from the interaction of excitons with the radial electric field and the potential gradient around the tip, both of which are current-dependent. At a given bias voltage, the electric field in the biased tip-sample junction increases with the tunneling current, because the tip-to-sample distance decreases when the current setpoint is increased. The radial potential gradient experienced by the excitons in the semiconductor also increases with the current. Indeed, at the positive sample biases at which STML is observed, electrons tunnel from the tip into the conduction band of the semiconductor, resulting in charging and lateral bending of the semiconductor electronic bands around the tip position~\cite{PenaRoman2022}. These effects become stronger with increasing tunneling current.

\subsubsection{Emission hotspots away from the excitation point}

%
\begin{figure*}
	\includegraphics[width=1.0\linewidth]{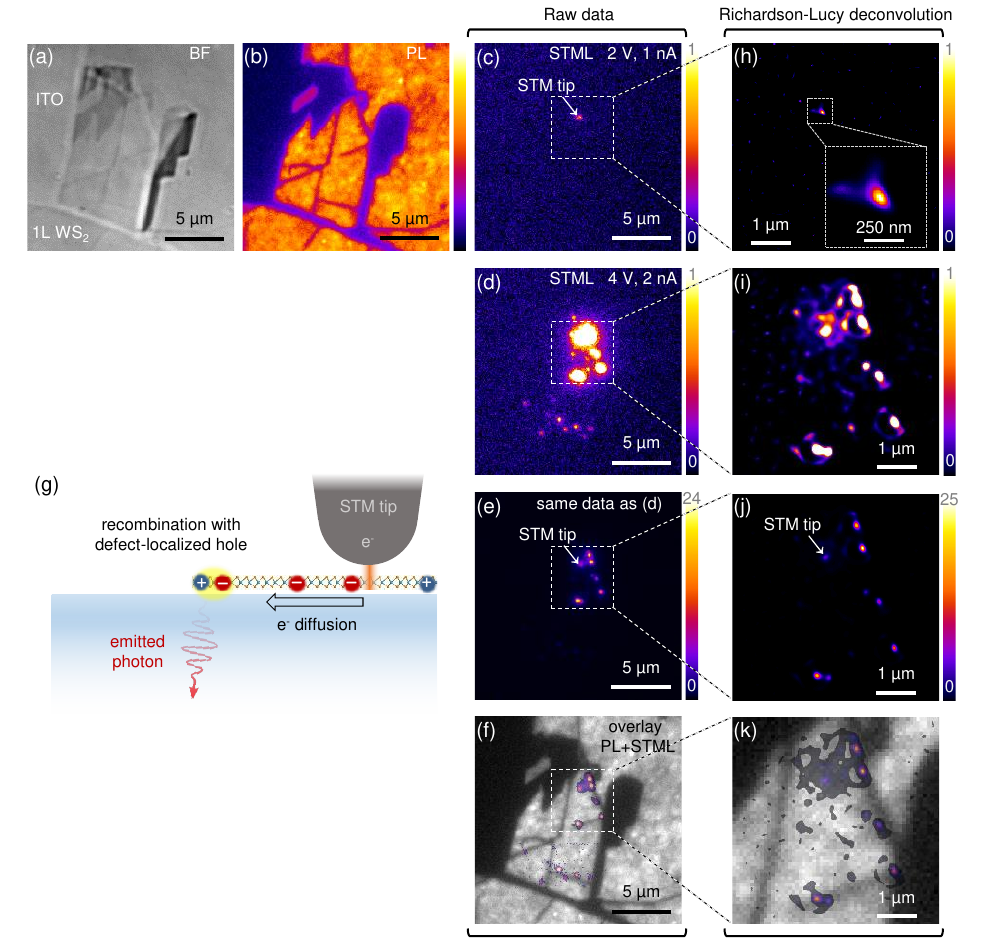} 
	\caption{(a) White-light transmission and (b) wide-field PL microscopy images of the sample: a flake of \ce{WS_2} exhibiting monolayer (1L) and multilayer areas. (c) to (e) Raw-data optical microscopy images of the STML measured on the same area as in (a) and (b). The lateral position of the STM tip is the same in (c) to (e). STM parameters: (c) $V_\mathrm{s}=2.0$~V, $I_\mathrm{t}=1$~nA; (d) $V_\mathrm{s}=4.0$~V, $I_\mathrm{t}=2$~nA. Acquisition time is $150$~s for both (c) and (d). For comparison purposes, the data is plotted on the same intensity scale in (c) and (d), yielding a saturated image in (d). The unsaturated image shown in (e) is the same data as in (d), plotted on a ($24$ times) larger intensity scale. (f) Overlay of the (grayscale) PL image shown in (b) and the STML data shown in (e). (g) Schematic of the local injection of electrons in monolayer \ce{WS_2} from the STM tip and their diffusion before they radiatively recombine with defect-trapped holes at flake folds or edges. (h) to (k) Images based on the same data as (c) to (f), respectively, except that the data are deconvolved using the Richardson-Lucy algorithm (script~2) on $10\times$ upsampled data. Moreover, a smaller area is considered in (h) to (k) as compared to (c) to (f), corresponding to the area delineated by a dotted square in images (c) to (f). In (h), a zoom of the STML spot is shown.}
	\label{FIG-3}
\end{figure*}

In the following, we again use deconvolution of real-space optical microscopy images to examine another little-studied and poorly understood effect of the STML of 2D semiconductors involving exciton dynamics. This time the effect may not be simply explained by exciton diffusion and drift away from the tunnel junction. In 2019, Pommier \textit{et al}~\cite{Pommier2019} reported the observation of STML emission hotspots on monolayer molybdenum diselenide (\ce{MoSe_2}), $1$ to $2~\mu$m away from the lateral position of the STM tip. This effect was attributed to the fact that excitons generated under the tip could diffuse over several micrometers before radiatively recombining, the hotspots being defects trapping the excitons. However, the results in Fig.~\ref{FIG-2} of the present work, obtained using RL deconvolution, show that typical exciton diffusion and drift distances in this type of STML experiments in ambient air are of the order of a hundred nanometers for the considered STM parameters. The reason why these hotspots were attributed to exciton diffusion is largely related to the misleading
effect of the optical microscope PSF, which, when convolved with the true distribution, leads to an apparently broadened spatial distribution of emitters. As shown in Fig.~\ref{FIG-1}(e), the optical microscope PSF is a peak function whose tail extends over more than about a micrometer before becoming negligible compared to its maximum intensity. After RL deconvolution, such a tail is no longer present in the spatial distribution of emitters [see Figs.~\ref{FIG-1}(e) and \ref{FIG-1}(f)]. The apparent spatial overlap of the tail of the emission spot under the tip and the distant hotspots is therefore an effect of convolution of the true distribution with the PSF and as such does not necessarily demonstrate delocalized emission over a micrometer-scale area due to exciton diffusion. Below, we show this unambiguously and propose an alternative explanation for these hotspots. Furthermore, it was not known until now whether the occurrence of STML emission hotspots was specific to a particular TMD, sample, or STM parameters. Such hotspots are not observed for monolayer \ce{WSe_2} in the image shown in Fig.~\ref{FIG-1}(d) and were not reported by Pe\~{n}a Rom\'{a}n \textit{et al} for monolayer \ce{WS_2} either~\cite{PenaRoman2022a}. Below, we show that STML emission hotspots may also be observed on monolayer \ce{WS_2} and that their occurrence depends on the STM parameters used.

Figures~\ref{FIG-3}(a) and~\ref{FIG-3}(b) show bright-field and PL microscopy images of a \ce{WS_2} flake, respectively. These two images are measured on the same area of the \ce{WS_2} flake. The bright areas in the PL image correspond to where the flake is monolayer in thickness (i.e., has a direct band gap) and the thin dark lines on the monolayer areas are likely due to nanofolds in the flake~\cite{Pommier2019} (formed during flake transfer). Figures~\ref{FIG-3}(c) to~\ref{FIG-3}(e) show optical microscopy images of the STML measured on the same area as in Figs.~\ref{FIG-3}(a) and~\ref{FIG-3}(b). The lateral position of the STM tip on a monolayer is the same in Figs.~\ref{FIG-3}(c) to~\ref{FIG-3}(e) and corresponds to the position of the bright spot indicated by a white arrow in Fig.~\ref{FIG-3}(c). The image shown in Fig.~\ref{FIG-3}(c) is measured at $V_\mathrm{s} = 2$~V and $I_\mathrm{t} = 1$~nA. Only one emission spot is seen, at the tip position, and no hot spots are observed. The image shown in Fig.~\ref{FIG-3}(d) is measured at $V_\mathrm{s} = 4$~V and $I_\mathrm{t} = 2$~nA and is plotted on the same intensity scale as the image shown in Fig.~\ref{FIG-3}(c). The same data but plotted on an unsaturated intensity scale are shown in Fig.~\ref{FIG-3}(e). A series of hot spots microns distant from the excitation source is seen, the most distant being located about $8.4~\mu$m from the tip position. The unsaturated image in Fig.~\ref{FIG-3}(e) shows that there is not a one-to-one relationship between the distance of the hot spots from the excitation source and their intensity. In addition, some hot spots are (up to $2.5$ times) more intense than the emission spot under the tip. These observations contradict the idea that the hot spots are the emission of excitons generated under the tip and propagating over several micrometers before radiatively recombining, since a decay-like distance dependence with an intensity maximum under the tip would be expected in this case. The superposition of the STML data and the PL image (in grayscale), shown in Fig.~\ref{FIG-3}(f), reveals that the most intense hot spots are close to nanofolds, corroborating the observations made by Pommier \textit{et al}~\cite{Pommier2019}. This also indicates that the largest observed distance separating the hot spots from the excitation source is not limited here by the exciton (or charge carrier) diffusion length, but by the geometry of the sample, i.e., the presence of nanofolds. 

Based on our observations, we propose a mechanism for STML emission hotspots which is schematized in Fig.~\ref{FIG-3}(g). As discussed above, the observed intensity and distance distribution of the hotspots with respect to the excitation source is inconsistent with a mechanism based on exciton diffusion. Therefore, we propose the following origin for the observed hot spots [see Fig.~\ref{FIG-3}(g)]: electrons injected from the tip into the semiconductor conduction band diffuse laterally away from the tip before recombining radiatively with holes localized at defects of the monolayer. This radiative recombination may occur via a defect-trapped exciton formation step at the defect sites. Monolayer TMDs deposited on ITO have been shown to be n-doped~\cite{Pommier2019,Morozov2021,PenaRoman2022,PenaRoman2022a,Brito2024}, which may explain why the injection of electrons into the conduction band does not activate the luminescence of the whole monolayer area, since holes are not available in the semiconductor valence band for the injected electrons to recombine with. However, defects may act as minority carrier (i.e., hole) traps. Moreover, the electronic transport properties of monolayer TMDs are known to be sensitive both to electronic band bending and external electric fields in the in-plane direction, the amplitude of which is current- and voltage-dependent in STM-based experiments. This may explain why the appearance of hot spots is only observed at sample biases as high as $4$~V and not at $2$~V for example, without this having any connection to the threshold bias voltage necessary for generating excitons (i.e., via inelastic tunneling)~\cite{PenaRoman2022a}. 

Finally, we consider an area of interest framed with a dashed line in Figs.~\ref{FIG-3}(c) to ~\ref{FIG-3}(e), and deconvolve it using the RL algorithm applied to $10 \times$-upsampled versions of the raw images. The measured STML spectrum of monolayer \ce{WS_2} is used to simulate the PSF~\cite{PenaRoman2022a}. From here on, we only use the second script for the RL algorithm (see Methods), which we have shown produces the fewest artifacts [see Fig.~\ref{FIG-1}]. The resulting deconvolved images are shown in Figs.~\ref{FIG-3}(h) to~\ref{FIG-3}(j). Zooming in on the emission spot in Fig.~\ref{FIG-3}(h) we can see more finely the spatial distribution of emitters; the central peak is of asymmetric shape and has a FWHM equal to approximately $0.105\pm0.005~\mu$m along the long axis and $0.065\pm0.005~\mu$m along the short axis and has one or more satellite spots of lower intensity. Expressed in terms of fractions of the emission wavelength ($\lambda_0=610$~nm)~\cite{PenaRoman2022a}, the latter values of FWHM correspond to $\lambda_0/5.8\pm0.3$ and $\lambda_0/9.4\pm0.7$, respectively. This is two to three times smaller than the Abbe diffraction limit $\lambda_0/2 \mathrm{NA} = \lambda_0/2.98 = 0.205~\mu$m at the considered emission wavelength. 

%
\begin{figure*}
	\includegraphics[width=1.0\linewidth]{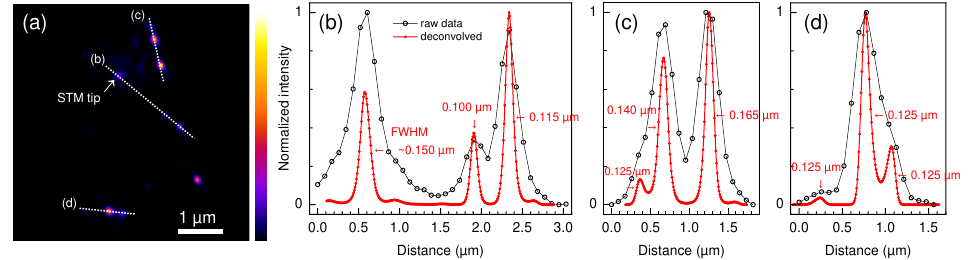} 
	\caption{(a) Optical microscopy image of STML measured on monolayer \ce{WS_2}, deconvolved using Richardson-Lucy algorithm on $10\times$ upsampled data [same data as in Fig.~\ref{FIG-3}(j)]. (b) to (d) Intensity profiles obtained along the dotted lines in the deconvolved image shown in (a) (red lines) and in the corresponding raw (not deconvolved, not upsampled) image shown in Fig.~\ref{FIG-3}(e) (black lines). The FWHM values shown in (b) to (d) are obtained from a 1D Gaussian fit of the presented profiles and are given to an accuracy of $0.005~\mu$m.}
	\label{FIG-4}
\end{figure*}

The deconvolution of the image recorded at $V_\mathrm{s} = 4$~V and the overlay with the PL image shown in Fig.~\ref{FIG-3}(k) confirm the presence of multiple hot spots, contained in an area of the monolayer delimited by the nearest nanofolds. The most intense hot spots appear aligned with the edges of the nanofolds. RL deconvolution makes it possible to distinguish many hot spots in Figs.~\ref{FIG-3}(i) and~\ref{FIG-3}(j), which are spatially unresolved in the raw image shown in Fig.~\ref{FIG-3}(e) due to the diffraction-limited resolution of the optical microscope. Figure~\ref{FIG-4} shows intensity profiles obtained from some of the STML emission spots of the deconvolved image shown in Fig.~\ref{FIG-3}(j) [also shown in Fig.~\ref{FIG-4}(a)]. The first intensity peak in Fig.~\ref{FIG-4}(b) corresponds to the emission spot under the STM tip, while the other peaks in Figs.~\ref{FIG-4}(b) to~\ref{FIG-4}(d) correspond to distant hot spots. Note that the two leftmost peaks in Fig.~\ref{FIG-4}(c) and the rightmost peaks in Fig.~\ref{FIG-4}(d), which are perfectly separated in the deconvolved data, are not resolved (i.e., distinguishable from each other) in the undeconvolved data. These peaks are separated by $0.295$ and $0.305\pm0.005~\mu$m, respectively, and the resolving power of the microscope, generally defined by the Rayleigh criterion $0.61\lambda_0/\mathrm{NA}$, is equal to $0.250~\mu$m here. These peaks are therefore not resolved without deconvolution because they are the image of spatial emitter distributions that have a finite width (while point emitters are assumed in the definition of the Rayleigh criterion). The effective spatial resolution achieved locally in the deconvolved image is therefore better than the FWHM of the emission spots in the deconvolved images.

Most of the hotspots visible in Fig.~\ref{FIG-4}(a) have an asymmetric shape, with those along a nanofold being elongated in the direction parallel to the nanofold. Therefore, profiles (b) and (c) in Fig.~\ref{FIG-4} obtained along nanofolds give the FWHM of the hotspots in the direction in which they are elongated. These FWHM values range from $0.100$ to $0.165~\mu$m (i.e., $\lambda_0/6.1$ and $\lambda_0/3.7$). In the direction orthogonal to the nanofold, the hotspots have a smaller FWHM, between $0.090$ and $0.105~\mu$m (i.e., between $\lambda_0/6.8$ and $\lambda_0/5.8$). 

\subsection{Final discussion}

The anisotropy discussed above cannot be unambiguously interpreted as an anisotropy in the spatial distribution of emitters at these nanofolds, because a single, uniform, and anisotropic spatial resolution (or effective PSF) cannot be defined for the entire deconvolved image. Since RL algorithms have been used in astronomy for decades, it is known that the effective spatial resolution achieved by these deconvolution methods for the image of two nearby emitters is lower along the axis crossing these two emitters than along the direction orthogonal to this axis~\cite{Lucy1992,Lucy1992a}. Moreover, the FWHM values extracted from the deconvolved images should be considered as upper limits to both the width of the emitter spatial distribution and the local effective spatial resolution. Here, due to the principle of the RL algorithms, the effective spatial resolution is limited by the noise level and the number of photons detected in the experiment, as well as by the convergence criteria of the algorithm. 

With these precautions in mind, the information on the spatial distribution of emitters around and away from the excitation source extracted from the RL deconvolved images is valuable for understanding the mechanisms of STML and, more broadly, the elementary electronic and excitonic processes involved in the operation of tunneling current-driven TMD-based light-emitting nanodevices. We anticipate that this type of information will be used in future studies for the development of models that better integrate the spatial aspect of these mechanisms and elementary processes. This type of information may only be obtained here because the STM excitation is combined with wide-field optical microscopy. The spatial distribution of emitters around the excitation source is simply not resolved in experiments based on photon mapping, i.e., scanning the STM tip over the sample and simultaneously recording the optical signal.

\section{Conclusions}

In conclusion, the spatial distribution of radiatively recombining excitons around the excitation source in the STML of \ce{WSe_2} and \ce{WS_2} monolayers was determined with sub-diffraction spatial resolution. This was achieved by combining STM with real-space optical microscopy and by applying an iterative image deconvolution algorithm, using an analytical dipole model of the optical microscope PSF. We showed that the width of the spatial distribution of emitters around the tip position increases with the current setpoint. We also demonstrated that the appearance of emission hotspots located a few micrometers from the excitation source is bias-dependent and likely results from electron diffusion and recombination with holes trapped by defects in the semiconductor. Such results may not be obtained in conventional STML experiments where the optical signal is acquired in an imaging mode based on STM tip scanning, i.e., using photon mapping. In future work, this technique could be used to study how exciton drift and diffusion properties depend on external electric fields and mechanical stresses, spatially heterogeneous doping (e.g., via multiple gate electrodes), or in lateral junctions of van der Waals materials, or hybrid heterostructures of nanoparticles or nanowires on 2D semiconductors.

\begin{acknowledgments}
This work is supported by public grants from the Agence Nationale de la Recherche (ATOEMS ANR-20-CE24-0010, TEXTURES ANR-22-CE09-0008, INFERNO ANR-22-CE42-0015, and EXODUS ANR-23-QUAC-0004). 
We are grateful to the STnano cleanroom staff for technical support. This work of the Interdisciplinary Thematic Institute QMat, as part of the ITI 2021–2028 program of the University of Strasbourg, CNRS and Inserm, was supported by IdEx Unistra (ANR 10 IDEX 0002), and by the SFRI STRAT'US project (ANR 20 SFRI 0012) and EUR QMAT ANR-17-EURE-0024 under the framework of the French ``Investments for the Future'' program. S.B. and A.R.M. acknowledge support from the Indo-French Centre for the Promotion of Advanced Research (CEFIPRA).
This work has received financial support from the Funda\c{c}\~{a}o de Amparo \`{a} Pesquisa do Estado de S\~{a}o Paulo (FAPESP), through projects 18/08543-7, 20/12480-0 and 14/23399-9. S.M. acknowledges support through the 2023 ``Optics in the City of Light'' NSF-REU program. The authors thank Guillaume Schull, Andrey G. Borisov, G\'{e}rald Dujardin and Sandrine L\'{e}v\^{e}que-Fort for discussions and encouragement.
\end{acknowledgments}


\begin{thebibliography}{85}%
	\makeatletter
	\providecommand \@ifxundefined [1]{%
		\@ifx{#1\undefined}
	}%
	\providecommand \@ifnum [1]{%
		\ifnum #1\expandafter \@firstoftwo
		\else \expandafter \@secondoftwo
		\fi
	}%
	\providecommand \@ifx [1]{%
		\ifx #1\expandafter \@firstoftwo
		\else \expandafter \@secondoftwo
		\fi
	}%
	\providecommand \natexlab [1]{#1}%
	\providecommand \enquote  [1]{``#1''}%
	\providecommand \bibnamefont  [1]{#1}%
	\providecommand \bibfnamefont [1]{#1}%
	\providecommand \citenamefont [1]{#1}%
	\providecommand \href@noop [0]{\@secondoftwo}%
	\providecommand \href [0]{\begingroup \@sanitize@url \@href}%
	\providecommand \@href[1]{\@@startlink{#1}\@@href}%
	\providecommand \@@href[1]{\endgroup#1\@@endlink}%
	\providecommand \@sanitize@url [0]{\catcode `\\12\catcode `\$12\catcode
		`\&12\catcode `\#12\catcode `\^12\catcode `\_12\catcode `\%12\relax}%
	\providecommand \@@startlink[1]{}%
	\providecommand \@@endlink[0]{}%
	\providecommand \url  [0]{\begingroup\@sanitize@url \@url }%
	\providecommand \@url [1]{\endgroup\@href {#1}{\urlprefix }}%
	\providecommand \urlprefix  [0]{URL }%
	\providecommand \Eprint [0]{\href }%
	\providecommand \doibase [0]{https://doi.org/}%
	\providecommand \selectlanguage [0]{\@gobble}%
	\providecommand \bibinfo  [0]{\@secondoftwo}%
	\providecommand \bibfield  [0]{\@secondoftwo}%
	\providecommand \translation [1]{[#1]}%
	\providecommand \BibitemOpen [0]{}%
	\providecommand \bibitemStop [0]{}%
	\providecommand \bibitemNoStop [0]{.\EOS\space}%
	\providecommand \EOS [0]{\spacefactor3000\relax}%
	\providecommand \BibitemShut  [1]{\csname bibitem#1\endcsname}%
	\let\auto@bib@innerbib\@empty
	\bibitem [{\citenamefont {Mak}\ \emph {et~al.}(2010)\citenamefont {Mak},
		\citenamefont {Lee}, \citenamefont {Hone}, \citenamefont {Shan},\ and\
		\citenamefont {Heinz}}]{Mak2010}%
	\BibitemOpen
	\bibfield  {author} {\bibinfo {author} {\bibfnamefont {K.~F.}\ \bibnamefont
			{Mak}}, \bibinfo {author} {\bibfnamefont {C.}~\bibnamefont {Lee}}, \bibinfo
		{author} {\bibfnamefont {J.}~\bibnamefont {Hone}}, \bibinfo {author}
		{\bibfnamefont {J.}~\bibnamefont {Shan}},\ and\ \bibinfo {author}
		{\bibfnamefont {T.~F.}\ \bibnamefont {Heinz}},\ }\bibfield  {title} {\bibinfo
		{title} {Atomically {Thin} \ce{MoS_2}: {A} {New} {Direct}-{Gap}
			{Semiconductor}},\ }\href {https://doi.org/10.1103/PhysRevLett.105.136805}
	{\bibfield  {journal} {\bibinfo  {journal} {Phys. Rev. Lett.}\ }\textbf
		{\bibinfo {volume} {105}},\ \bibinfo {pages} {136805} (\bibinfo {year}
		{2010})}\BibitemShut {NoStop}%
	\bibitem [{\citenamefont {Splendiani}\ \emph {et~al.}(2010)\citenamefont
		{Splendiani}, \citenamefont {Sun}, \citenamefont {Zhang}, \citenamefont {Li},
		\citenamefont {Kim}, \citenamefont {Chim}, \citenamefont {Galli},\ and\
		\citenamefont {Wang}}]{Splendiani2010}%
	\BibitemOpen
	\bibfield  {author} {\bibinfo {author} {\bibfnamefont {A.}~\bibnamefont
			{Splendiani}}, \bibinfo {author} {\bibfnamefont {L.}~\bibnamefont {Sun}},
		\bibinfo {author} {\bibfnamefont {Y.}~\bibnamefont {Zhang}}, \bibinfo
		{author} {\bibfnamefont {T.}~\bibnamefont {Li}}, \bibinfo {author}
		{\bibfnamefont {J.}~\bibnamefont {Kim}}, \bibinfo {author} {\bibfnamefont
			{C.-Y.}\ \bibnamefont {Chim}}, \bibinfo {author} {\bibfnamefont
			{G.}~\bibnamefont {Galli}},\ and\ \bibinfo {author} {\bibfnamefont
			{F.}~\bibnamefont {Wang}},\ }\bibfield  {title} {\bibinfo {title} {Emerging
			{Photoluminescence} in {Monolayer} \ce{MoS_2}},\ }\href
	{https://doi.org/10.1021/nl903868w} {\bibfield  {journal} {\bibinfo
			{journal} {Nano Lett.}\ }\textbf {\bibinfo {volume} {10}},\ \bibinfo {pages}
		{1271} (\bibinfo {year} {2010})}\BibitemShut {NoStop}%
	\bibitem [{\citenamefont {Geim}\ and\ \citenamefont
		{Grigorieva}(2013)}]{Geim2013}%
	\BibitemOpen
	\bibfield  {author} {\bibinfo {author} {\bibfnamefont {A.~K.}\ \bibnamefont
			{Geim}}\ and\ \bibinfo {author} {\bibfnamefont {I.~V.}\ \bibnamefont
			{Grigorieva}},\ }\bibfield  {title} {\bibinfo {title} {Van der {Waals}
			heterostructures},\ }\href {https://doi.org/10.1038/nature12385} {\bibfield
		{journal} {\bibinfo  {journal} {Nature}\ }\textbf {\bibinfo {volume} {499}},\
		\bibinfo {pages} {419} (\bibinfo {year} {2013})}\BibitemShut {NoStop}%
	\bibitem [{\citenamefont {Novoselov}\ \emph {et~al.}(2016)\citenamefont
		{Novoselov}, \citenamefont {Mishchenko}, \citenamefont {Carvalho},\ and\
		\citenamefont {Neto}}]{Novoselov2016}%
	\BibitemOpen
	\bibfield  {author} {\bibinfo {author} {\bibfnamefont {K.~S.}\ \bibnamefont
			{Novoselov}}, \bibinfo {author} {\bibfnamefont {A.}~\bibnamefont
			{Mishchenko}}, \bibinfo {author} {\bibfnamefont {A.}~\bibnamefont
			{Carvalho}},\ and\ \bibinfo {author} {\bibfnamefont {A.~H.~C.}\ \bibnamefont
			{Neto}},\ }\bibfield  {title} {\bibinfo {title} {{2D} materials and van der
			waals heterostructures},\ }\href {https://doi.org/10.1126/science.aac9439}
	{\bibfield  {journal} {\bibinfo  {journal} {Science}\ }\textbf {\bibinfo
			{volume} {353}},\ \bibinfo {pages} {aac9439} (\bibinfo {year}
		{2016})}\BibitemShut {NoStop}%
	\bibitem [{\citenamefont {Brar}\ \emph {et~al.}(2017)\citenamefont {Brar},
		\citenamefont {Koltonow},\ and\ \citenamefont {Huang}}]{Brar2017}%
	\BibitemOpen
	\bibfield  {author} {\bibinfo {author} {\bibfnamefont {V.~W.}\ \bibnamefont
			{Brar}}, \bibinfo {author} {\bibfnamefont {A.~R.}\ \bibnamefont {Koltonow}},\
		and\ \bibinfo {author} {\bibfnamefont {J.}~\bibnamefont {Huang}},\ }\bibfield
	{title} {\bibinfo {title} {New {Discoveries} and {Opportunities} from
			{Two}-{Dimensional} {Materials}},\ }\href
	{https://doi.org/10.1021/acsphotonics.7b00194} {\bibfield  {journal}
		{\bibinfo  {journal} {ACS Photonics}\ }\textbf {\bibinfo {volume} {4}},\
		\bibinfo {pages} {407} (\bibinfo {year} {2017})}\BibitemShut {NoStop}%
	\bibitem [{\citenamefont {Binder}\ \emph {et~al.}(2017)\citenamefont {Binder},
		\citenamefont {Withers}, \citenamefont {Molas}, \citenamefont {Faugeras},
		\citenamefont {Nogajewski}, \citenamefont {Watanabe}, \citenamefont
		{Taniguchi}, \citenamefont {Kozikov}, \citenamefont {Geim}, \citenamefont
		{Novoselov},\ and\ \citenamefont {Potemski}}]{Binder2017}%
	\BibitemOpen
	\bibfield  {author} {\bibinfo {author} {\bibfnamefont {J.}~\bibnamefont
			{Binder}}, \bibinfo {author} {\bibfnamefont {F.}~\bibnamefont {Withers}},
		\bibinfo {author} {\bibfnamefont {M.~R.}\ \bibnamefont {Molas}}, \bibinfo
		{author} {\bibfnamefont {C.}~\bibnamefont {Faugeras}}, \bibinfo {author}
		{\bibfnamefont {K.}~\bibnamefont {Nogajewski}}, \bibinfo {author}
		{\bibfnamefont {K.}~\bibnamefont {Watanabe}}, \bibinfo {author}
		{\bibfnamefont {T.}~\bibnamefont {Taniguchi}}, \bibinfo {author}
		{\bibfnamefont {A.}~\bibnamefont {Kozikov}}, \bibinfo {author} {\bibfnamefont
			{A.~K.}\ \bibnamefont {Geim}}, \bibinfo {author} {\bibfnamefont {K.~S.}\
			\bibnamefont {Novoselov}},\ and\ \bibinfo {author} {\bibfnamefont
			{M.}~\bibnamefont {Potemski}},\ }\bibfield  {title} {\bibinfo {title}
		{Sub-bandgap voltage electroluminescence and magneto-oscillations in a
			\ce{WSe_2} light-emitting van der waals heterostructure},\ }\href@noop {}
	{\bibfield  {journal} {\bibinfo  {journal} {Nano Lett.}\ }\textbf {\bibinfo
			{volume} {17}},\ \bibinfo {pages} {1425} (\bibinfo {year}
		{2017})}\BibitemShut {NoStop}%
	\bibitem [{\citenamefont {Branny}\ \emph {et~al.}(2017)\citenamefont {Branny},
		\citenamefont {Kumar}, \citenamefont {Proux},\ and\ \citenamefont
		{Gerardot}}]{Branny2017}%
	\BibitemOpen
	\bibfield  {author} {\bibinfo {author} {\bibfnamefont {A.}~\bibnamefont
			{Branny}}, \bibinfo {author} {\bibfnamefont {S.}~\bibnamefont {Kumar}},
		\bibinfo {author} {\bibfnamefont {R.}~\bibnamefont {Proux}},\ and\ \bibinfo
		{author} {\bibfnamefont {B.~D.}\ \bibnamefont {Gerardot}},\ }\bibfield
	{title} {\bibinfo {title} {Deterministic strain-induced arrays of quantum
			emitters in a two-dimensional semiconductor},\ }\href
	{https://doi.org/10.1038/ncomms15053} {\bibfield  {journal} {\bibinfo
			{journal} {Nat. Commun.}\ }\textbf {\bibinfo {volume} {8}},\ \bibinfo {pages}
		{15053} (\bibinfo {year} {2017})}\BibitemShut {NoStop}%
	\bibitem [{\citenamefont {Xie}\ \emph {et~al.}(2018)\citenamefont {Xie},
		\citenamefont {Tu}, \citenamefont {Han}, \citenamefont {Huang}, \citenamefont
		{Kang}, \citenamefont {Lao}, \citenamefont {Poddar}, \citenamefont {Park},
		\citenamefont {Muller}, \citenamefont {DiStasio},\ and\ \citenamefont
		{Park}}]{Xie2018}%
	\BibitemOpen
	\bibfield  {author} {\bibinfo {author} {\bibfnamefont {S.}~\bibnamefont
			{Xie}}, \bibinfo {author} {\bibfnamefont {L.}~\bibnamefont {Tu}}, \bibinfo
		{author} {\bibfnamefont {Y.}~\bibnamefont {Han}}, \bibinfo {author}
		{\bibfnamefont {L.}~\bibnamefont {Huang}}, \bibinfo {author} {\bibfnamefont
			{K.}~\bibnamefont {Kang}}, \bibinfo {author} {\bibfnamefont {K.~U.}\
			\bibnamefont {Lao}}, \bibinfo {author} {\bibfnamefont {P.}~\bibnamefont
			{Poddar}}, \bibinfo {author} {\bibfnamefont {C.}~\bibnamefont {Park}},
		\bibinfo {author} {\bibfnamefont {D.~A.}\ \bibnamefont {Muller}}, \bibinfo
		{author} {\bibfnamefont {R.~A.}\ \bibnamefont {DiStasio}},\ and\ \bibinfo
		{author} {\bibfnamefont {J.}~\bibnamefont {Park}},\ }\bibfield  {title}
	{\bibinfo {title} {Coherent, atomically thin transition-metal dichalcogenide
			superlattices with engineered strain},\ }\href
	{https://doi.org/10.1126/science.aao5360} {\bibfield  {journal} {\bibinfo
			{journal} {Science}\ }\textbf {\bibinfo {volume} {359}},\ \bibinfo {pages}
		{1131} (\bibinfo {year} {2018})}\BibitemShut {NoStop}%
	\bibitem [{\citenamefont {Qi}\ \emph {et~al.}(2023)\citenamefont {Qi},
		\citenamefont {Sadi}, \citenamefont {Hu}, \citenamefont {Zheng},
		\citenamefont {Wu}, \citenamefont {Jiang},\ and\ \citenamefont
		{Chen}}]{Qi2023}%
	\BibitemOpen
	\bibfield  {author} {\bibinfo {author} {\bibfnamefont {Y.}~\bibnamefont
			{Qi}}, \bibinfo {author} {\bibfnamefont {M.~A.}\ \bibnamefont {Sadi}},
		\bibinfo {author} {\bibfnamefont {D.}~\bibnamefont {Hu}}, \bibinfo {author}
		{\bibfnamefont {M.}~\bibnamefont {Zheng}}, \bibinfo {author} {\bibfnamefont
			{Z.}~\bibnamefont {Wu}}, \bibinfo {author} {\bibfnamefont {Y.}~\bibnamefont
			{Jiang}},\ and\ \bibinfo {author} {\bibfnamefont {Y.~P.}\ \bibnamefont
			{Chen}},\ }\bibfield  {title} {\bibinfo {title} {Recent progress in strain
			engineering on van der waals {2D} materials: Tunable electrical,
			electrochemical, magnetic, and optical properties},\ }\href
	{https://doi.org/10.1002/adma.202205714} {\bibfield  {journal} {\bibinfo
			{journal} {Adv. Mater.}\ }\textbf {\bibinfo {volume} {35}},\ \bibinfo {pages}
		{2205714} (\bibinfo {year} {2023})}\BibitemShut {NoStop}%
	\bibitem [{\citenamefont {Anichini}\ \emph {et~al.}(2018)\citenamefont
		{Anichini}, \citenamefont {Czepa}, \citenamefont {Pakulski}, \citenamefont
		{Aliprandi}, \citenamefont {Ciesielski},\ and\ \citenamefont
		{Samor\`{i}}}]{Anichini2018}%
	\BibitemOpen
	\bibfield  {author} {\bibinfo {author} {\bibfnamefont {C.}~\bibnamefont
			{Anichini}}, \bibinfo {author} {\bibfnamefont {W.}~\bibnamefont {Czepa}},
		\bibinfo {author} {\bibfnamefont {D.}~\bibnamefont {Pakulski}}, \bibinfo
		{author} {\bibfnamefont {A.}~\bibnamefont {Aliprandi}}, \bibinfo {author}
		{\bibfnamefont {A.}~\bibnamefont {Ciesielski}},\ and\ \bibinfo {author}
		{\bibfnamefont {P.}~\bibnamefont {Samor\`{i}}},\ }\bibfield  {title}
	{\bibinfo {title} {Chemical sensing with {2D} materials},\ }\href
	{https://doi.org/10.1039/C8CS00417J} {\bibfield  {journal} {\bibinfo
			{journal} {Chem. Soc. Rev.}\ }\textbf {\bibinfo {volume} {47}},\ \bibinfo
		{pages} {4860} (\bibinfo {year} {2018})}\BibitemShut {NoStop}%
	\bibitem [{\citenamefont {Meng}\ \emph {et~al.}(2019)\citenamefont {Meng},
		\citenamefont {Stolz}, \citenamefont {Mendecki},\ and\ \citenamefont
		{Mirica}}]{Meng2019}%
	\BibitemOpen
	\bibfield  {author} {\bibinfo {author} {\bibfnamefont {Z.}~\bibnamefont
			{Meng}}, \bibinfo {author} {\bibfnamefont {R.~M.}\ \bibnamefont {Stolz}},
		\bibinfo {author} {\bibfnamefont {L.}~\bibnamefont {Mendecki}},\ and\
		\bibinfo {author} {\bibfnamefont {K.~A.}\ \bibnamefont {Mirica}},\ }\bibfield
	{title} {\bibinfo {title} {Electrically-transduced chemical sensors based on
			two-dimensional nanomaterials},\ }\href
	{https://doi.org/10.1021/acs.chemrev.8b00311} {\bibfield  {journal} {\bibinfo
			{journal} {Chem. Rev.}\ }\textbf {\bibinfo {volume} {119}},\ \bibinfo
		{pages} {478} (\bibinfo {year} {2019})}\BibitemShut {NoStop}%
	\bibitem [{\citenamefont {Pang}\ \emph {et~al.}(2020)\citenamefont {Pang},
		\citenamefont {Yang}, \citenamefont {Yang},\ and\ \citenamefont
		{Ren}}]{Pang2020}%
	\BibitemOpen
	\bibfield  {author} {\bibinfo {author} {\bibfnamefont {Y.}~\bibnamefont
			{Pang}}, \bibinfo {author} {\bibfnamefont {Z.}~\bibnamefont {Yang}}, \bibinfo
		{author} {\bibfnamefont {Y.}~\bibnamefont {Yang}},\ and\ \bibinfo {author}
		{\bibfnamefont {T.-L.}\ \bibnamefont {Ren}},\ }\bibfield  {title} {\bibinfo
		{title} {Wearable electronics based on {2D} materials for human physiological
			information detection},\ }\href {https://doi.org/10.1002/smll.201901124}
	{\bibfield  {journal} {\bibinfo  {journal} {Small}\ }\textbf {\bibinfo
			{volume} {16}},\ \bibinfo {pages} {1901124} (\bibinfo {year}
		{2020})}\BibitemShut {NoStop}%
	\bibitem [{\citenamefont {Khan}\ \emph {et~al.}(2021)\citenamefont {Khan},
		\citenamefont {Tareen}, \citenamefont {Wang}, \citenamefont {Aslam},
		\citenamefont {Ma}, \citenamefont {Mahmood}, \citenamefont {Ouyang},
		\citenamefont {Zhang},\ and\ \citenamefont {Guo}}]{Khan2021}%
	\BibitemOpen
	\bibfield  {author} {\bibinfo {author} {\bibfnamefont {K.}~\bibnamefont
			{Khan}}, \bibinfo {author} {\bibfnamefont {A.~K.}\ \bibnamefont {Tareen}},
		\bibinfo {author} {\bibfnamefont {L.}~\bibnamefont {Wang}}, \bibinfo {author}
		{\bibfnamefont {M.}~\bibnamefont {Aslam}}, \bibinfo {author} {\bibfnamefont
			{C.}~\bibnamefont {Ma}}, \bibinfo {author} {\bibfnamefont {N.}~\bibnamefont
			{Mahmood}}, \bibinfo {author} {\bibfnamefont {Z.}~\bibnamefont {Ouyang}},
		\bibinfo {author} {\bibfnamefont {H.}~\bibnamefont {Zhang}},\ and\ \bibinfo
		{author} {\bibfnamefont {Z.}~\bibnamefont {Guo}},\ }\bibfield  {title}
	{\bibinfo {title} {Sensing applications of atomically thin group iv carbon
			siblings xenes: Progress, challenges, and prospects},\ }\href
	{https://doi.org/10.1002/adfm.202005957} {\bibfield  {journal} {\bibinfo
			{journal} {Adv. Funct. Mater.}\ }\textbf {\bibinfo {volume} {31}},\ \bibinfo
		{pages} {2005957} (\bibinfo {year} {2021})}\BibitemShut {NoStop}%
	\bibitem [{\citenamefont {Rohaizad}\ \emph {et~al.}(2021)\citenamefont
		{Rohaizad}, \citenamefont {Mayorga-Martinez}, \citenamefont {Fojt\r{u}},
		\citenamefont {Latiff},\ and\ \citenamefont {Pumera}}]{Rohaizad2021}%
	\BibitemOpen
	\bibfield  {author} {\bibinfo {author} {\bibfnamefont {N.}~\bibnamefont
			{Rohaizad}}, \bibinfo {author} {\bibfnamefont {C.~C.}\ \bibnamefont
			{Mayorga-Martinez}}, \bibinfo {author} {\bibfnamefont {M.}~\bibnamefont
			{Fojt\r{u}}}, \bibinfo {author} {\bibfnamefont {N.~M.}\ \bibnamefont
			{Latiff}},\ and\ \bibinfo {author} {\bibfnamefont {M.}~\bibnamefont
			{Pumera}},\ }\bibfield  {title} {\bibinfo {title} {Two-dimensional materials
			in biomedical{,} biosensing and sensing applications},\ }\href
	{https://doi.org/10.1039/D0CS00150C} {\bibfield  {journal} {\bibinfo
			{journal} {Chem. Soc. Rev.}\ }\textbf {\bibinfo {volume} {50}},\ \bibinfo
		{pages} {619} (\bibinfo {year} {2021})}\BibitemShut {NoStop}%
	\bibitem [{\citenamefont {Wang}\ \emph {et~al.}(2012)\citenamefont {Wang},
		\citenamefont {Kalantar-Zadeh}, \citenamefont {Kis}, \citenamefont
		{Coleman},\ and\ \citenamefont {Strano}}]{Wang2012}%
	\BibitemOpen
	\bibfield  {author} {\bibinfo {author} {\bibfnamefont {Q.~H.}\ \bibnamefont
			{Wang}}, \bibinfo {author} {\bibfnamefont {K.}~\bibnamefont
			{Kalantar-Zadeh}}, \bibinfo {author} {\bibfnamefont {A.}~\bibnamefont {Kis}},
		\bibinfo {author} {\bibfnamefont {J.~N.}\ \bibnamefont {Coleman}},\ and\
		\bibinfo {author} {\bibfnamefont {M.~S.}\ \bibnamefont {Strano}},\ }\bibfield
	{title} {\bibinfo {title} {Electronics and optoelectronics of
			two-dimensional transition metal dichalcogenides},\ }\href
	{https://doi.org/10.1038/nnano.2012.193} {\bibfield  {journal} {\bibinfo
			{journal} {Nat. Nanotechnol.}\ }\textbf {\bibinfo {volume} {7}},\ \bibinfo
		{pages} {699} (\bibinfo {year} {2012})}\BibitemShut {NoStop}%
	\bibitem [{\citenamefont {Cheng}\ \emph {et~al.}(2014)\citenamefont {Cheng},
		\citenamefont {Li}, \citenamefont {Zhou}, \citenamefont {Wang}, \citenamefont
		{Yin}, \citenamefont {Jiang}, \citenamefont {Liu}, \citenamefont {Chen},
		\citenamefont {Huang},\ and\ \citenamefont {Duan}}]{Cheng2014}%
	\BibitemOpen
	\bibfield  {author} {\bibinfo {author} {\bibfnamefont {R.}~\bibnamefont
			{Cheng}}, \bibinfo {author} {\bibfnamefont {D.}~\bibnamefont {Li}}, \bibinfo
		{author} {\bibfnamefont {H.}~\bibnamefont {Zhou}}, \bibinfo {author}
		{\bibfnamefont {C.}~\bibnamefont {Wang}}, \bibinfo {author} {\bibfnamefont
			{A.}~\bibnamefont {Yin}}, \bibinfo {author} {\bibfnamefont {S.}~\bibnamefont
			{Jiang}}, \bibinfo {author} {\bibfnamefont {Y.}~\bibnamefont {Liu}}, \bibinfo
		{author} {\bibfnamefont {Y.}~\bibnamefont {Chen}}, \bibinfo {author}
		{\bibfnamefont {Y.}~\bibnamefont {Huang}},\ and\ \bibinfo {author}
		{\bibfnamefont {X.}~\bibnamefont {Duan}},\ }\bibfield  {title} {\bibinfo
		{title} {Electroluminescence and {Photocurrent} {Generation} from
			{Atomically} {Sharp} \ce{WSe_2}/\ce{MoS_2} {Heterojunction} p-n {Diodes}},\
	}\href {https://doi.org/10.1021/nl502075n} {\bibfield  {journal} {\bibinfo
			{journal} {Nano Lett.}\ }\textbf {\bibinfo {volume} {14}},\ \bibinfo {pages}
		{5590} (\bibinfo {year} {2014})}\BibitemShut {NoStop}%
	\bibitem [{\citenamefont {Furchi}\ \emph {et~al.}(2014)\citenamefont {Furchi},
		\citenamefont {Pospischil}, \citenamefont {Libisch}, \citenamefont
		{Burgd\"{o}rfer},\ and\ \citenamefont {Mueller}}]{Furchi2014}%
	\BibitemOpen
	\bibfield  {author} {\bibinfo {author} {\bibfnamefont {M.~M.}\ \bibnamefont
			{Furchi}}, \bibinfo {author} {\bibfnamefont {A.}~\bibnamefont {Pospischil}},
		\bibinfo {author} {\bibfnamefont {F.}~\bibnamefont {Libisch}}, \bibinfo
		{author} {\bibfnamefont {J.}~\bibnamefont {Burgd\"{o}rfer}},\ and\ \bibinfo
		{author} {\bibfnamefont {T.}~\bibnamefont {Mueller}},\ }\bibfield  {title}
	{\bibinfo {title} {Photovoltaic {Effect} in an {Electrically} {Tunable} van
			der {Waals} {Heterojunction}},\ }\href {https://doi.org/10.1021/nl501962c}
	{\bibfield  {journal} {\bibinfo  {journal} {Nano Lett.}\ }\textbf {\bibinfo
			{volume} {14}},\ \bibinfo {pages} {4785} (\bibinfo {year}
		{2014})}\BibitemShut {NoStop}%
	\bibitem [{\citenamefont {Ross}\ \emph {et~al.}(2014)\citenamefont {Ross},
		\citenamefont {Klement}, \citenamefont {Jones}, \citenamefont {Ghimire},
		\citenamefont {Yan}, \citenamefont {Mandrus}, \citenamefont {Taniguchi},
		\citenamefont {Watanabe}, \citenamefont {Kitamura}, \citenamefont {Yao},
		\citenamefont {Cobden},\ and\ \citenamefont {Xu}}]{Ross2014}%
	\BibitemOpen
	\bibfield  {author} {\bibinfo {author} {\bibfnamefont {J.~S.}\ \bibnamefont
			{Ross}}, \bibinfo {author} {\bibfnamefont {P.}~\bibnamefont {Klement}},
		\bibinfo {author} {\bibfnamefont {A.~M.}\ \bibnamefont {Jones}}, \bibinfo
		{author} {\bibfnamefont {N.~J.}\ \bibnamefont {Ghimire}}, \bibinfo {author}
		{\bibfnamefont {J.}~\bibnamefont {Yan}}, \bibinfo {author} {\bibfnamefont
			{D.~G.}\ \bibnamefont {Mandrus}}, \bibinfo {author} {\bibfnamefont
			{T.}~\bibnamefont {Taniguchi}}, \bibinfo {author} {\bibfnamefont
			{K.}~\bibnamefont {Watanabe}}, \bibinfo {author} {\bibfnamefont
			{K.}~\bibnamefont {Kitamura}}, \bibinfo {author} {\bibfnamefont
			{W.}~\bibnamefont {Yao}}, \bibinfo {author} {\bibfnamefont {D.~H.}\
			\bibnamefont {Cobden}},\ and\ \bibinfo {author} {\bibfnamefont
			{X.}~\bibnamefont {Xu}},\ }\bibfield  {title} {\bibinfo {title} {Electrically
			tunable excitonic light-emitting diodes based on monolayer \ce{WSe_2} p-n
			junctions},\ }\href {https://doi.org/10.1038/nnano.2014.26} {\bibfield
		{journal} {\bibinfo  {journal} {Nat. Nanotechnol.}\ }\textbf {\bibinfo
			{volume} {9}},\ \bibinfo {pages} {268} (\bibinfo {year} {2014})}\BibitemShut
	{NoStop}%
	\bibitem [{\citenamefont {Akinwande}\ \emph {et~al.}(2014)\citenamefont
		{Akinwande}, \citenamefont {Petrone},\ and\ \citenamefont
		{Hone}}]{Akinwande2014}%
	\BibitemOpen
	\bibfield  {author} {\bibinfo {author} {\bibfnamefont {D.}~\bibnamefont
			{Akinwande}}, \bibinfo {author} {\bibfnamefont {N.}~\bibnamefont {Petrone}},\
		and\ \bibinfo {author} {\bibfnamefont {J.}~\bibnamefont {Hone}},\ }\bibfield
	{title} {\bibinfo {title} {Two-dimensional flexible nanoelectronics},\ }\href
	{https://doi.org/10.1038/ncomms6678} {\bibfield  {journal} {\bibinfo
			{journal} {Nat. Commun.}\ }\textbf {\bibinfo {volume} {5}},\ \bibinfo {pages}
		{5678} (\bibinfo {year} {2014})}\BibitemShut {NoStop}%
	\bibitem [{\citenamefont {Clark}\ \emph {et~al.}(2016)\citenamefont {Clark},
		\citenamefont {Schaibley}, \citenamefont {Ross}, \citenamefont {Taniguchi},
		\citenamefont {Watanabe}, \citenamefont {Hendrickson}, \citenamefont {Mou},
		\citenamefont {Yao},\ and\ \citenamefont {Xu}}]{Clark2016}%
	\BibitemOpen
	\bibfield  {author} {\bibinfo {author} {\bibfnamefont {G.}~\bibnamefont
			{Clark}}, \bibinfo {author} {\bibfnamefont {J.~R.}\ \bibnamefont
			{Schaibley}}, \bibinfo {author} {\bibfnamefont {J.}~\bibnamefont {Ross}},
		\bibinfo {author} {\bibfnamefont {T.}~\bibnamefont {Taniguchi}}, \bibinfo
		{author} {\bibfnamefont {K.}~\bibnamefont {Watanabe}}, \bibinfo {author}
		{\bibfnamefont {J.~R.}\ \bibnamefont {Hendrickson}}, \bibinfo {author}
		{\bibfnamefont {S.}~\bibnamefont {Mou}}, \bibinfo {author} {\bibfnamefont
			{W.}~\bibnamefont {Yao}},\ and\ \bibinfo {author} {\bibfnamefont
			{X.}~\bibnamefont {Xu}},\ }\bibfield  {title} {\bibinfo {title} {Single
			{Defect} {Light}-{Emitting} {Diode} in a van der {Waals} {Heterostructure}},\
	}\href {https://doi.org/10.1021/acs.nanolett.6b01580} {\bibfield  {journal}
		{\bibinfo  {journal} {Nano Lett.}\ }\textbf {\bibinfo {volume} {16}},\
		\bibinfo {pages} {3944} (\bibinfo {year} {2016})}\BibitemShut {NoStop}%
	\bibitem [{\citenamefont {Jariwala}\ \emph {et~al.}(2016)\citenamefont
		{Jariwala}, \citenamefont {Howell}, \citenamefont {Chen}, \citenamefont
		{Kang}, \citenamefont {Sangwan}, \citenamefont {Filippone}, \citenamefont
		{Turrisi}, \citenamefont {Marks}, \citenamefont {Lauhon},\ and\ \citenamefont
		{Hersam}}]{Jariwala2016}%
	\BibitemOpen
	\bibfield  {author} {\bibinfo {author} {\bibfnamefont {D.}~\bibnamefont
			{Jariwala}}, \bibinfo {author} {\bibfnamefont {S.~L.}\ \bibnamefont
			{Howell}}, \bibinfo {author} {\bibfnamefont {K.-S.}\ \bibnamefont {Chen}},
		\bibinfo {author} {\bibfnamefont {J.}~\bibnamefont {Kang}}, \bibinfo {author}
		{\bibfnamefont {V.~K.}\ \bibnamefont {Sangwan}}, \bibinfo {author}
		{\bibfnamefont {S.~A.}\ \bibnamefont {Filippone}}, \bibinfo {author}
		{\bibfnamefont {R.}~\bibnamefont {Turrisi}}, \bibinfo {author} {\bibfnamefont
			{T.~J.}\ \bibnamefont {Marks}}, \bibinfo {author} {\bibfnamefont {L.~J.}\
			\bibnamefont {Lauhon}},\ and\ \bibinfo {author} {\bibfnamefont {M.~C.}\
			\bibnamefont {Hersam}},\ }\bibfield  {title} {\bibinfo {title} {Hybrid,
			{Gate}-{Tunable}, van der {Waals} p-n {Heterojunctions} from {Pentacene} and
			\ce{MoS_2}},\ }\href {https://doi.org/10.1021/acs.nanolett.5b04141}
	{\bibfield  {journal} {\bibinfo  {journal} {Nano Lett.}\ }\textbf {\bibinfo
			{volume} {16}},\ \bibinfo {pages} {497} (\bibinfo {year} {2016})}\BibitemShut
	{NoStop}%
	\bibitem [{\citenamefont {Mak}\ and\ \citenamefont {Shan}(2016)}]{Mak2016}%
	\BibitemOpen
	\bibfield  {author} {\bibinfo {author} {\bibfnamefont {K.~F.}\ \bibnamefont
			{Mak}}\ and\ \bibinfo {author} {\bibfnamefont {J.}~\bibnamefont {Shan}},\
	}\bibfield  {title} {\bibinfo {title} {Photonics and optoelectronics of {2D}
			semiconductor transition metal dichalcogenides},\ }\href
	{https://doi.org/10.1038/nphoton.2015.282} {\bibfield  {journal} {\bibinfo
			{journal} {Nat. Photonics}\ }\textbf {\bibinfo {volume} {10}},\ \bibinfo
		{pages} {216} (\bibinfo {year} {2016})}\BibitemShut {NoStop}%
	\bibitem [{\citenamefont {Liu}\ \emph {et~al.}(2017)\citenamefont {Liu},
		\citenamefont {Clark}, \citenamefont {Fryett}, \citenamefont {Wu},
		\citenamefont {Zheng}, \citenamefont {Hatami}, \citenamefont {Xu},\ and\
		\citenamefont {Majumdar}}]{Liu2017}%
	\BibitemOpen
	\bibfield  {author} {\bibinfo {author} {\bibfnamefont {C.-H.}\ \bibnamefont
			{Liu}}, \bibinfo {author} {\bibfnamefont {G.}~\bibnamefont {Clark}}, \bibinfo
		{author} {\bibfnamefont {T.}~\bibnamefont {Fryett}}, \bibinfo {author}
		{\bibfnamefont {S.}~\bibnamefont {Wu}}, \bibinfo {author} {\bibfnamefont
			{J.}~\bibnamefont {Zheng}}, \bibinfo {author} {\bibfnamefont
			{F.}~\bibnamefont {Hatami}}, \bibinfo {author} {\bibfnamefont
			{X.}~\bibnamefont {Xu}},\ and\ \bibinfo {author} {\bibfnamefont
			{A.}~\bibnamefont {Majumdar}},\ }\bibfield  {title} {\bibinfo {title}
		{Nanocavity {Integrated} van der {Waals} {Heterostructure} {Light}-{Emitting}
			{Tunneling} {Diode}},\ }\href {https://doi.org/10.1021/acs.nanolett.6b03801}
	{\bibfield  {journal} {\bibinfo  {journal} {Nano Lett.}\ }\textbf {\bibinfo
			{volume} {17}},\ \bibinfo {pages} {200} (\bibinfo {year} {2017})}\BibitemShut
	{NoStop}%
	\bibitem [{\citenamefont {Brar}\ \emph {et~al.}(2018)\citenamefont {Brar},
		\citenamefont {Sherrott},\ and\ \citenamefont {Jariwala}}]{Brar2018}%
	\BibitemOpen
	\bibfield  {author} {\bibinfo {author} {\bibfnamefont {V.~W.}\ \bibnamefont
			{Brar}}, \bibinfo {author} {\bibfnamefont {M.~C.}\ \bibnamefont {Sherrott}},\
		and\ \bibinfo {author} {\bibfnamefont {D.}~\bibnamefont {Jariwala}},\
	}\bibfield  {title} {\bibinfo {title} {Emerging photonic architectures in
			two-dimensional opto-electronics},\ }\href
	{https://doi.org/10.1039/C8CS00206A} {\bibfield  {journal} {\bibinfo
			{journal} {Chem. Soc. Rev.}\ }\textbf {\bibinfo {volume} {47}},\ \bibinfo
		{pages} {6824} (\bibinfo {year} {2018})}\BibitemShut {NoStop}%
	\bibitem [{\citenamefont {Lemme}\ \emph {et~al.}(2022)\citenamefont {Lemme},
		\citenamefont {Akinwande}, \citenamefont {Huyghebaert},\ and\ \citenamefont
		{Stampfer}}]{Lemme2022}%
	\BibitemOpen
	\bibfield  {author} {\bibinfo {author} {\bibfnamefont {M.~C.}\ \bibnamefont
			{Lemme}}, \bibinfo {author} {\bibfnamefont {D.}~\bibnamefont {Akinwande}},
		\bibinfo {author} {\bibfnamefont {C.}~\bibnamefont {Huyghebaert}},\ and\
		\bibinfo {author} {\bibfnamefont {C.}~\bibnamefont {Stampfer}},\ }\bibfield
	{title} {\bibinfo {title} {{{2D}} materials for future heterogeneous
			electronics},\ }\href {https://doi.org/10.1038/s41467-022-29001-4} {\bibfield
		{journal} {\bibinfo  {journal} {Nat. Commun.}\ }\textbf {\bibinfo {volume}
			{13}},\ \bibinfo {pages} {1392} (\bibinfo {year} {2022})}\BibitemShut
	{NoStop}%
	\bibitem [{\citenamefont {Palacios-Berraquero}\ \emph
		{et~al.}(2017)\citenamefont {Palacios-Berraquero}, \citenamefont {Kara},
		\citenamefont {Montblanch}, \citenamefont {Barbone}, \citenamefont
		{Latawiec}, \citenamefont {Yoon}, \citenamefont {Ott}, \citenamefont
		{Loncar}, \citenamefont {Ferrari},\ and\ \citenamefont
		{Atat\"{u}re}}]{Palacios-Berraquero2017}%
	\BibitemOpen
	\bibfield  {author} {\bibinfo {author} {\bibfnamefont {C.}~\bibnamefont
			{Palacios-Berraquero}}, \bibinfo {author} {\bibfnamefont {D.~M.}\
			\bibnamefont {Kara}}, \bibinfo {author} {\bibfnamefont {A.~R.-P.}\
			\bibnamefont {Montblanch}}, \bibinfo {author} {\bibfnamefont
			{M.}~\bibnamefont {Barbone}}, \bibinfo {author} {\bibfnamefont
			{P.}~\bibnamefont {Latawiec}}, \bibinfo {author} {\bibfnamefont
			{D.}~\bibnamefont {Yoon}}, \bibinfo {author} {\bibfnamefont {A.~K.}\
			\bibnamefont {Ott}}, \bibinfo {author} {\bibfnamefont {M.}~\bibnamefont
			{Loncar}}, \bibinfo {author} {\bibfnamefont {A.~C.}\ \bibnamefont
			{Ferrari}},\ and\ \bibinfo {author} {\bibfnamefont {M.}~\bibnamefont
			{Atat\"{u}re}},\ }\bibfield  {title} {\bibinfo {title} {Large-scale
			quantum-emitter arrays in atomically thin semiconductors},\ }\href
	{https://doi.org/10.1038/ncomms15093} {\bibfield  {journal} {\bibinfo
			{journal} {Nat. Commun.}\ }\textbf {\bibinfo {volume} {8}},\ \bibinfo {pages}
		{15093} (\bibinfo {year} {2017})}\BibitemShut {NoStop}%
	\bibitem [{\citenamefont {Raja}\ \emph {et~al.}(2017)\citenamefont {Raja},
		\citenamefont {Chaves}, \citenamefont {Yu}, \citenamefont {Arefe},
		\citenamefont {Hill}, \citenamefont {Rigosi}, \citenamefont {Berkelbach},
		\citenamefont {Nagler}, \citenamefont {Schüller}, \citenamefont {Korn},
		\citenamefont {Nuckolls}, \citenamefont {Hone}, \citenamefont {Brus},
		\citenamefont {Heinz}, \citenamefont {Reichman},\ and\ \citenamefont
		{Chernikov}}]{Raja2017}%
	\BibitemOpen
	\bibfield  {author} {\bibinfo {author} {\bibfnamefont {A.}~\bibnamefont
			{Raja}}, \bibinfo {author} {\bibfnamefont {A.}~\bibnamefont {Chaves}},
		\bibinfo {author} {\bibfnamefont {J.}~\bibnamefont {Yu}}, \bibinfo {author}
		{\bibfnamefont {G.}~\bibnamefont {Arefe}}, \bibinfo {author} {\bibfnamefont
			{H.~M.}\ \bibnamefont {Hill}}, \bibinfo {author} {\bibfnamefont {A.~F.}\
			\bibnamefont {Rigosi}}, \bibinfo {author} {\bibfnamefont {T.~C.}\
			\bibnamefont {Berkelbach}}, \bibinfo {author} {\bibfnamefont
			{P.}~\bibnamefont {Nagler}}, \bibinfo {author} {\bibfnamefont
			{C.}~\bibnamefont {Schüller}}, \bibinfo {author} {\bibfnamefont
			{T.}~\bibnamefont {Korn}}, \bibinfo {author} {\bibfnamefont {C.}~\bibnamefont
			{Nuckolls}}, \bibinfo {author} {\bibfnamefont {J.}~\bibnamefont {Hone}},
		\bibinfo {author} {\bibfnamefont {L.~E.}\ \bibnamefont {Brus}}, \bibinfo
		{author} {\bibfnamefont {T.~F.}\ \bibnamefont {Heinz}}, \bibinfo {author}
		{\bibfnamefont {D.~R.}\ \bibnamefont {Reichman}},\ and\ \bibinfo {author}
		{\bibfnamefont {A.}~\bibnamefont {Chernikov}},\ }\bibfield  {title} {\bibinfo
		{title} {Coulomb engineering of the bandgap and excitons in two-dimensional
			materials},\ }\href {https://doi.org/10.1038/ncomms15251} {\bibfield
		{journal} {\bibinfo  {journal} {Nat. Commun.}\ }\textbf {\bibinfo {volume}
			{8}},\ \bibinfo {pages} {15251} (\bibinfo {year} {2017})}\BibitemShut
	{NoStop}%
	\bibitem [{\citenamefont {Wang}\ \emph {et~al.}(2018)\citenamefont {Wang},
		\citenamefont {Chernikov}, \citenamefont {Glazov}, \citenamefont {Heinz},
		\citenamefont {Marie}, \citenamefont {Amand},\ and\ \citenamefont
		{Urbaszek}}]{Wang2018}%
	\BibitemOpen
	\bibfield  {author} {\bibinfo {author} {\bibfnamefont {G.}~\bibnamefont
			{Wang}}, \bibinfo {author} {\bibfnamefont {A.}~\bibnamefont {Chernikov}},
		\bibinfo {author} {\bibfnamefont {M.~M.}\ \bibnamefont {Glazov}}, \bibinfo
		{author} {\bibfnamefont {T.~F.}\ \bibnamefont {Heinz}}, \bibinfo {author}
		{\bibfnamefont {X.}~\bibnamefont {Marie}}, \bibinfo {author} {\bibfnamefont
			{T.}~\bibnamefont {Amand}},\ and\ \bibinfo {author} {\bibfnamefont
			{B.}~\bibnamefont {Urbaszek}},\ }\bibfield  {title} {\bibinfo {title}
		{Colloquium: Excitons in atomically thin transition metal dichalcogenides},\
	}\href {https://doi.org/10.1103/RevModPhys.90.021001} {\bibfield  {journal}
		{\bibinfo  {journal} {Rev. Mod. Phys.}\ }\textbf {\bibinfo {volume} {90}},\
		\bibinfo {pages} {021001} (\bibinfo {year} {2018})}\BibitemShut {NoStop}%
	\bibitem [{\citenamefont {Unuchek}\ \emph {et~al.}(2018)\citenamefont
		{Unuchek}, \citenamefont {Ciarrocchi}, \citenamefont {Avsar}, \citenamefont
		{Watanabe}, \citenamefont {Taniguchi},\ and\ \citenamefont
		{Kis}}]{Unuchek2018}%
	\BibitemOpen
	\bibfield  {author} {\bibinfo {author} {\bibfnamefont {D.}~\bibnamefont
			{Unuchek}}, \bibinfo {author} {\bibfnamefont {A.}~\bibnamefont {Ciarrocchi}},
		\bibinfo {author} {\bibfnamefont {A.}~\bibnamefont {Avsar}}, \bibinfo
		{author} {\bibfnamefont {K.}~\bibnamefont {Watanabe}}, \bibinfo {author}
		{\bibfnamefont {T.}~\bibnamefont {Taniguchi}},\ and\ \bibinfo {author}
		{\bibfnamefont {A.}~\bibnamefont {Kis}},\ }\bibfield  {title} {\bibinfo
		{title} {Room-temperature electrical control of exciton flux in a van der
			{Waals} heterostructure},\ }\href {https://doi.org/10.1038/s41586-018-0357-y}
	{\bibfield  {journal} {\bibinfo  {journal} {Nature}\ }\textbf {\bibinfo
			{volume} {560}},\ \bibinfo {pages} {340} (\bibinfo {year}
		{2018})}\BibitemShut {NoStop}%
	\bibitem [{\citenamefont {Jauregui}\ \emph {et~al.}(2019)\citenamefont
		{Jauregui}, \citenamefont {Joe}, \citenamefont {Pistunova}, \citenamefont
		{Wild}, \citenamefont {High}, \citenamefont {Zhou}, \citenamefont {Scuri},
		\citenamefont {Greve}, \citenamefont {Sushko}, \citenamefont {Yu},
		\citenamefont {Taniguchi}, \citenamefont {Watanabe}, \citenamefont
		{Needleman}, \citenamefont {Lukin}, \citenamefont {Park},\ and\ \citenamefont
		{Kim}}]{Jauregui2019}%
	\BibitemOpen
	\bibfield  {author} {\bibinfo {author} {\bibfnamefont {L.~A.}\ \bibnamefont
			{Jauregui}}, \bibinfo {author} {\bibfnamefont {A.~Y.}\ \bibnamefont {Joe}},
		\bibinfo {author} {\bibfnamefont {K.}~\bibnamefont {Pistunova}}, \bibinfo
		{author} {\bibfnamefont {D.~S.}\ \bibnamefont {Wild}}, \bibinfo {author}
		{\bibfnamefont {A.~A.}\ \bibnamefont {High}}, \bibinfo {author}
		{\bibfnamefont {Y.}~\bibnamefont {Zhou}}, \bibinfo {author} {\bibfnamefont
			{G.}~\bibnamefont {Scuri}}, \bibinfo {author} {\bibfnamefont {K.~D.}\
			\bibnamefont {Greve}}, \bibinfo {author} {\bibfnamefont {A.}~\bibnamefont
			{Sushko}}, \bibinfo {author} {\bibfnamefont {C.-H.}\ \bibnamefont {Yu}},
		\bibinfo {author} {\bibfnamefont {T.}~\bibnamefont {Taniguchi}}, \bibinfo
		{author} {\bibfnamefont {K.}~\bibnamefont {Watanabe}}, \bibinfo {author}
		{\bibfnamefont {D.~J.}\ \bibnamefont {Needleman}}, \bibinfo {author}
		{\bibfnamefont {M.~D.}\ \bibnamefont {Lukin}}, \bibinfo {author}
		{\bibfnamefont {H.}~\bibnamefont {Park}},\ and\ \bibinfo {author}
		{\bibfnamefont {P.}~\bibnamefont {Kim}},\ }\bibfield  {title} {\bibinfo
		{title} {Electrical control of interlayer exciton dynamics in atomically thin
			heterostructures},\ }\href {https://doi.org/10.1126/science.aaw4194}
	{\bibfield  {journal} {\bibinfo  {journal} {Science}\ }\textbf {\bibinfo
			{volume} {366}},\ \bibinfo {pages} {870} (\bibinfo {year}
		{2019})}\BibitemShut {NoStop}%
	\bibitem [{\citenamefont {Baek}\ \emph {et~al.}(2020)\citenamefont {Baek},
		\citenamefont {Brotons-Gisbert}, \citenamefont {Koong}, \citenamefont
		{Campbell}, \citenamefont {Rambach}, \citenamefont {Watanabe}, \citenamefont
		{Taniguchi},\ and\ \citenamefont {Gerardot}}]{Baek2020}%
	\BibitemOpen
	\bibfield  {author} {\bibinfo {author} {\bibfnamefont {H.}~\bibnamefont
			{Baek}}, \bibinfo {author} {\bibfnamefont {M.}~\bibnamefont
			{Brotons-Gisbert}}, \bibinfo {author} {\bibfnamefont {Z.~X.}\ \bibnamefont
			{Koong}}, \bibinfo {author} {\bibfnamefont {A.}~\bibnamefont {Campbell}},
		\bibinfo {author} {\bibfnamefont {M.}~\bibnamefont {Rambach}}, \bibinfo
		{author} {\bibfnamefont {K.}~\bibnamefont {Watanabe}}, \bibinfo {author}
		{\bibfnamefont {T.}~\bibnamefont {Taniguchi}},\ and\ \bibinfo {author}
		{\bibfnamefont {B.~D.}\ \bibnamefont {Gerardot}},\ }\bibfield  {title}
	{\bibinfo {title} {Highly energy-tunable quantum light from moir\'{e}-trapped
			excitons},\ }\href {https://doi.org/10.1126/sciadv.aba8526} {\bibfield
		{journal} {\bibinfo  {journal} {Sci. Adv.}\ }\textbf {\bibinfo {volume}
			{6}},\ \bibinfo {pages} {eaba8526} (\bibinfo {year} {2020})}\BibitemShut
	{NoStop}%
	\bibitem [{\citenamefont {Perea-Causin}\ \emph {et~al.}(2022)\citenamefont
		{Perea-Causin}, \citenamefont {Erkensten}, \citenamefont {Fitzgerald},
		\citenamefont {Thompson}, \citenamefont {Rosati}, \citenamefont {Brem},\ and\
		\citenamefont {Malic}}]{PereaCausin2022}%
	\BibitemOpen
	\bibfield  {author} {\bibinfo {author} {\bibfnamefont {R.}~\bibnamefont
			{Perea-Causin}}, \bibinfo {author} {\bibfnamefont {D.}~\bibnamefont
			{Erkensten}}, \bibinfo {author} {\bibfnamefont {J.~M.}\ \bibnamefont
			{Fitzgerald}}, \bibinfo {author} {\bibfnamefont {J.~J.~P.}\ \bibnamefont
			{Thompson}}, \bibinfo {author} {\bibfnamefont {R.}~\bibnamefont {Rosati}},
		\bibinfo {author} {\bibfnamefont {S.}~\bibnamefont {Brem}},\ and\ \bibinfo
		{author} {\bibfnamefont {E.}~\bibnamefont {Malic}},\ }\bibfield  {title}
	{\bibinfo {title} {Exciton optics, dynamics, and transport in atomically thin
			semiconductors},\ }\href {https://doi.org/10.1063/5.0107665} {\bibfield
		{journal} {\bibinfo  {journal} {APL Mater.}\ }\textbf {\bibinfo {volume}
			{10}},\ \bibinfo {pages} {100701} (\bibinfo {year} {2022})}\BibitemShut
	{NoStop}%
	\bibitem [{\citenamefont {Tagarelli}\ \emph {et~al.}(2023)\citenamefont
		{Tagarelli}, \citenamefont {Lopriore}, \citenamefont {Erkensten},
		\citenamefont {Perea-Caus\'{i}n}, \citenamefont {Brem}, \citenamefont
		{Hagel}, \citenamefont {Sun}, \citenamefont {Pasquale}, \citenamefont
		{Watanabe}, \citenamefont {Taniguchi}, \citenamefont {Malic},\ and\
		\citenamefont {Kis}}]{Tagarelli2023}%
	\BibitemOpen
	\bibfield  {author} {\bibinfo {author} {\bibfnamefont {F.}~\bibnamefont
			{Tagarelli}}, \bibinfo {author} {\bibfnamefont {E.}~\bibnamefont {Lopriore}},
		\bibinfo {author} {\bibfnamefont {D.}~\bibnamefont {Erkensten}}, \bibinfo
		{author} {\bibfnamefont {R.}~\bibnamefont {Perea-Caus\'{i}n}}, \bibinfo
		{author} {\bibfnamefont {S.}~\bibnamefont {Brem}}, \bibinfo {author}
		{\bibfnamefont {J.}~\bibnamefont {Hagel}}, \bibinfo {author} {\bibfnamefont
			{Z.}~\bibnamefont {Sun}}, \bibinfo {author} {\bibfnamefont {G.}~\bibnamefont
			{Pasquale}}, \bibinfo {author} {\bibfnamefont {K.}~\bibnamefont {Watanabe}},
		\bibinfo {author} {\bibfnamefont {T.}~\bibnamefont {Taniguchi}}, \bibinfo
		{author} {\bibfnamefont {E.}~\bibnamefont {Malic}},\ and\ \bibinfo {author}
		{\bibfnamefont {A.}~\bibnamefont {Kis}},\ }\bibfield  {title} {\bibinfo
		{title} {Electrical control of hybrid exciton transport in a van der {Waals}
			heterostructure},\ }\href {https://doi.org/10.1038/s41566-023-01198-w}
	{\bibfield  {journal} {\bibinfo  {journal} {Nat. Photonics}\ }\textbf
		{\bibinfo {volume} {17}},\ \bibinfo {pages} {615} (\bibinfo {year}
		{2023})}\BibitemShut {NoStop}%
	\bibitem [{\citenamefont {Malic}\ \emph {et~al.}(2023)\citenamefont {Malic},
		\citenamefont {Perea-Causin}, \citenamefont {Rosati}, \citenamefont
		{Erkensten},\ and\ \citenamefont {Brem}}]{Malic2023}%
	\BibitemOpen
	\bibfield  {author} {\bibinfo {author} {\bibfnamefont {E.}~\bibnamefont
			{Malic}}, \bibinfo {author} {\bibfnamefont {R.}~\bibnamefont {Perea-Causin}},
		\bibinfo {author} {\bibfnamefont {R.}~\bibnamefont {Rosati}}, \bibinfo
		{author} {\bibfnamefont {D.}~\bibnamefont {Erkensten}},\ and\ \bibinfo
		{author} {\bibfnamefont {S.}~\bibnamefont {Brem}},\ }\bibfield  {title}
	{\bibinfo {title} {Exciton transport in atomically thin semiconductors},\
	}\href {https://doi.org/10.1038/s41467-023-38556-9} {\bibfield  {journal}
		{\bibinfo  {journal} {Nat. Commun.}\ }\textbf {\bibinfo {volume} {14}},\
		\bibinfo {pages} {3430} (\bibinfo {year} {2023})}\BibitemShut {NoStop}%
	\bibitem [{\citenamefont {Koo}\ \emph {et~al.}(2024)\citenamefont {Koo},
		\citenamefont {Moon}, \citenamefont {Kang}, \citenamefont {Joo},
		\citenamefont {Lee}, \citenamefont {Lee}, \citenamefont {Kravtsov},\ and\
		\citenamefont {Park}}]{Koo2024}%
	\BibitemOpen
	\bibfield  {author} {\bibinfo {author} {\bibfnamefont {Y.}~\bibnamefont
			{Koo}}, \bibinfo {author} {\bibfnamefont {T.}~\bibnamefont {Moon}}, \bibinfo
		{author} {\bibfnamefont {M.}~\bibnamefont {Kang}}, \bibinfo {author}
		{\bibfnamefont {H.}~\bibnamefont {Joo}}, \bibinfo {author} {\bibfnamefont
			{C.}~\bibnamefont {Lee}}, \bibinfo {author} {\bibfnamefont {H.}~\bibnamefont
			{Lee}}, \bibinfo {author} {\bibfnamefont {V.}~\bibnamefont {Kravtsov}},\ and\
		\bibinfo {author} {\bibfnamefont {K.-D.}\ \bibnamefont {Park}},\ }\bibfield
	{title} {\bibinfo {title} {Dynamical control of nanoscale light-matter
			interactions in low-dimensional quantum materials},\ }\href
	{https://doi.org/10.1038/s41377-024-01380-x} {\bibfield  {journal} {\bibinfo
			{journal} {Light Sci. Appl.}\ }\textbf {\bibinfo {volume} {13}},\ \bibinfo
		{pages} {30} (\bibinfo {year} {2024})}\BibitemShut {NoStop}%
	\bibitem [{\citenamefont {Park}\ \emph {et~al.}(2016)\citenamefont {Park},
		\citenamefont {Khatib}, \citenamefont {Kravtsov}, \citenamefont {Clark},
		\citenamefont {Xu},\ and\ \citenamefont {Raschke}}]{Park2016}%
	\BibitemOpen
	\bibfield  {author} {\bibinfo {author} {\bibfnamefont {K.-D.}\ \bibnamefont
			{Park}}, \bibinfo {author} {\bibfnamefont {O.}~\bibnamefont {Khatib}},
		\bibinfo {author} {\bibfnamefont {V.}~\bibnamefont {Kravtsov}}, \bibinfo
		{author} {\bibfnamefont {G.}~\bibnamefont {Clark}}, \bibinfo {author}
		{\bibfnamefont {X.}~\bibnamefont {Xu}},\ and\ \bibinfo {author}
		{\bibfnamefont {M.~B.}\ \bibnamefont {Raschke}},\ }\bibfield  {title}
	{\bibinfo {title} {Hybrid {Tip}-{Enhanced} {Nanospectroscopy} and
			{Nanoimaging} of {Monolayer} \ce{WSe_2} with {Local} {Strain} {Control}},\
	}\href {https://doi.org/10.1021/acs.nanolett.6b00238} {\bibfield  {journal}
		{\bibinfo  {journal} {Nano Lett.}\ }\textbf {\bibinfo {volume} {16}},\
		\bibinfo {pages} {2621} (\bibinfo {year} {2016})}\BibitemShut {NoStop}%
	\bibitem [{\citenamefont {Park}\ \emph {et~al.}(2018)\citenamefont {Park},
		\citenamefont {Jiang}, \citenamefont {Clark}, \citenamefont {Xu},\ and\
		\citenamefont {Raschke}}]{Park2018}%
	\BibitemOpen
	\bibfield  {author} {\bibinfo {author} {\bibfnamefont {K.-D.}\ \bibnamefont
			{Park}}, \bibinfo {author} {\bibfnamefont {T.}~\bibnamefont {Jiang}},
		\bibinfo {author} {\bibfnamefont {G.}~\bibnamefont {Clark}}, \bibinfo
		{author} {\bibfnamefont {X.}~\bibnamefont {Xu}},\ and\ \bibinfo {author}
		{\bibfnamefont {M.~B.}\ \bibnamefont {Raschke}},\ }\bibfield  {title}
	{\bibinfo {title} {Radiative control of dark excitons at room temperature by
			nano-optical antenna-tip {Purcell} effect},\ }\href
	{https://doi.org/10.1038/s41565-017-0003-0} {\bibfield  {journal} {\bibinfo
			{journal} {Nat. Nanotechnol.}\ }\textbf {\bibinfo {volume} {13}},\ \bibinfo
		{pages} {59} (\bibinfo {year} {2018})}\BibitemShut {NoStop}%
	\bibitem [{\citenamefont {He}\ \emph {et~al.}(2019)\citenamefont {He},
		\citenamefont {Han}, \citenamefont {Yuan}, \citenamefont {Sinyukov},
		\citenamefont {Eleuch}, \citenamefont {Niu}, \citenamefont {Zhang},
		\citenamefont {Lou}, \citenamefont {Hu}, \citenamefont {Voronine},\ and\
		\citenamefont {Scully}}]{He2019}%
	\BibitemOpen
	\bibfield  {author} {\bibinfo {author} {\bibfnamefont {Z.}~\bibnamefont
			{He}}, \bibinfo {author} {\bibfnamefont {Z.}~\bibnamefont {Han}}, \bibinfo
		{author} {\bibfnamefont {J.}~\bibnamefont {Yuan}}, \bibinfo {author}
		{\bibfnamefont {A.~M.}\ \bibnamefont {Sinyukov}}, \bibinfo {author}
		{\bibfnamefont {H.}~\bibnamefont {Eleuch}}, \bibinfo {author} {\bibfnamefont
			{C.}~\bibnamefont {Niu}}, \bibinfo {author} {\bibfnamefont {Z.}~\bibnamefont
			{Zhang}}, \bibinfo {author} {\bibfnamefont {J.}~\bibnamefont {Lou}}, \bibinfo
		{author} {\bibfnamefont {J.}~\bibnamefont {Hu}}, \bibinfo {author}
		{\bibfnamefont {D.~V.}\ \bibnamefont {Voronine}},\ and\ \bibinfo {author}
		{\bibfnamefont {M.~O.}\ \bibnamefont {Scully}},\ }\bibfield  {title}
	{\bibinfo {title} {Quantum plasmonic control of trions in a picocavity with
			monolayer \ce{WS_2}},\ }\href {https://doi.org/10.1126/sciadv.aau8763}
	{\bibfield  {journal} {\bibinfo  {journal} {Sci. Adv.}\ }\textbf {\bibinfo
			{volume} {5}},\ \bibinfo {pages} {eaau8763} (\bibinfo {year}
		{2019})}\BibitemShut {NoStop}%
	\bibitem [{\citenamefont {Darlington}\ \emph {et~al.}(2020)\citenamefont
		{Darlington}, \citenamefont {Carmesin}, \citenamefont {Florian},
		\citenamefont {Yanev}, \citenamefont {Ajayi}, \citenamefont {Ardelean},
		\citenamefont {Rhodes}, \citenamefont {Ghiotto}, \citenamefont {Krayev},
		\citenamefont {Watanabe}, \citenamefont {Taniguchi}, \citenamefont {Kysar},
		\citenamefont {Pasupathy}, \citenamefont {Hone}, \citenamefont {Jahnke},
		\citenamefont {Borys},\ and\ \citenamefont {Schuck}}]{Darlington2020}%
	\BibitemOpen
	\bibfield  {author} {\bibinfo {author} {\bibfnamefont {T.~P.}\ \bibnamefont
			{Darlington}}, \bibinfo {author} {\bibfnamefont {C.}~\bibnamefont
			{Carmesin}}, \bibinfo {author} {\bibfnamefont {M.}~\bibnamefont {Florian}},
		\bibinfo {author} {\bibfnamefont {E.}~\bibnamefont {Yanev}}, \bibinfo
		{author} {\bibfnamefont {O.}~\bibnamefont {Ajayi}}, \bibinfo {author}
		{\bibfnamefont {J.}~\bibnamefont {Ardelean}}, \bibinfo {author}
		{\bibfnamefont {D.~A.}\ \bibnamefont {Rhodes}}, \bibinfo {author}
		{\bibfnamefont {A.}~\bibnamefont {Ghiotto}}, \bibinfo {author} {\bibfnamefont
			{A.}~\bibnamefont {Krayev}}, \bibinfo {author} {\bibfnamefont
			{K.}~\bibnamefont {Watanabe}}, \bibinfo {author} {\bibfnamefont
			{T.}~\bibnamefont {Taniguchi}}, \bibinfo {author} {\bibfnamefont {J.~W.}\
			\bibnamefont {Kysar}}, \bibinfo {author} {\bibfnamefont {A.~N.}\ \bibnamefont
			{Pasupathy}}, \bibinfo {author} {\bibfnamefont {J.~C.}\ \bibnamefont {Hone}},
		\bibinfo {author} {\bibfnamefont {F.}~\bibnamefont {Jahnke}}, \bibinfo
		{author} {\bibfnamefont {N.~J.}\ \bibnamefont {Borys}},\ and\ \bibinfo
		{author} {\bibfnamefont {P.~J.}\ \bibnamefont {Schuck}},\ }\bibfield  {title}
	{\bibinfo {title} {Imaging strain-localized excitons in nanoscale bubbles of
			monolayer {WSe2} at room temperature},\ }\href
	{https://doi.org/10.1038/s41565-020-0730-5} {\bibfield  {journal} {\bibinfo
			{journal} {Nat. Nanotechnol.}\ }\textbf {\bibinfo {volume} {15}},\ \bibinfo
		{pages} {854} (\bibinfo {year} {2020})}\BibitemShut {NoStop}%
	\bibitem [{\citenamefont {Zhang}\ \emph {et~al.}(2022)\citenamefont {Zhang},
		\citenamefont {Li}, \citenamefont {Chen}, \citenamefont {Ruta}, \citenamefont
		{Shao}, \citenamefont {Sternbach}, \citenamefont {McLeod}, \citenamefont
		{Sun}, \citenamefont {Xiong}, \citenamefont {Moore}, \citenamefont {Xu},
		\citenamefont {Wu}, \citenamefont {Shabani}, \citenamefont {Zhou},
		\citenamefont {Wang}, \citenamefont {Mooshammer}, \citenamefont {Ray},
		\citenamefont {Wilson}, \citenamefont {Schuck}, \citenamefont {Dean},
		\citenamefont {Pasupathy}, \citenamefont {Lipson}, \citenamefont {Xu},
		\citenamefont {Zhu}, \citenamefont {Millis}, \citenamefont {Liu},
		\citenamefont {Hone},\ and\ \citenamefont {Basov}}]{Zhang2022}%
	\BibitemOpen
	\bibfield  {author} {\bibinfo {author} {\bibfnamefont {S.}~\bibnamefont
			{Zhang}}, \bibinfo {author} {\bibfnamefont {B.}~\bibnamefont {Li}}, \bibinfo
		{author} {\bibfnamefont {X.}~\bibnamefont {Chen}}, \bibinfo {author}
		{\bibfnamefont {F.~L.}\ \bibnamefont {Ruta}}, \bibinfo {author}
		{\bibfnamefont {Y.}~\bibnamefont {Shao}}, \bibinfo {author} {\bibfnamefont
			{A.~J.}\ \bibnamefont {Sternbach}}, \bibinfo {author} {\bibfnamefont {A.~S.}\
			\bibnamefont {McLeod}}, \bibinfo {author} {\bibfnamefont {Z.}~\bibnamefont
			{Sun}}, \bibinfo {author} {\bibfnamefont {L.}~\bibnamefont {Xiong}}, \bibinfo
		{author} {\bibfnamefont {S.~L.}\ \bibnamefont {Moore}}, \bibinfo {author}
		{\bibfnamefont {X.}~\bibnamefont {Xu}}, \bibinfo {author} {\bibfnamefont
			{W.}~\bibnamefont {Wu}}, \bibinfo {author} {\bibfnamefont {S.}~\bibnamefont
			{Shabani}}, \bibinfo {author} {\bibfnamefont {L.}~\bibnamefont {Zhou}},
		\bibinfo {author} {\bibfnamefont {Z.}~\bibnamefont {Wang}}, \bibinfo {author}
		{\bibfnamefont {F.}~\bibnamefont {Mooshammer}}, \bibinfo {author}
		{\bibfnamefont {E.}~\bibnamefont {Ray}}, \bibinfo {author} {\bibfnamefont
			{N.}~\bibnamefont {Wilson}}, \bibinfo {author} {\bibfnamefont {P.~J.}\
			\bibnamefont {Schuck}}, \bibinfo {author} {\bibfnamefont {C.~R.}\
			\bibnamefont {Dean}}, \bibinfo {author} {\bibfnamefont {A.~N.}\ \bibnamefont
			{Pasupathy}}, \bibinfo {author} {\bibfnamefont {M.}~\bibnamefont {Lipson}},
		\bibinfo {author} {\bibfnamefont {X.}~\bibnamefont {Xu}}, \bibinfo {author}
		{\bibfnamefont {X.}~\bibnamefont {Zhu}}, \bibinfo {author} {\bibfnamefont
			{A.~J.}\ \bibnamefont {Millis}}, \bibinfo {author} {\bibfnamefont
			{M.}~\bibnamefont {Liu}}, \bibinfo {author} {\bibfnamefont {J.~C.}\
			\bibnamefont {Hone}},\ and\ \bibinfo {author} {\bibfnamefont {D.~N.}\
			\bibnamefont {Basov}},\ }\bibfield  {title} {\bibinfo {title}
		{Nano-spectroscopy of excitons in atomically thin transition metal
			dichalcogenides},\ }\href {https://doi.org/10.1038/s41467-022-28117-x}
	{\bibfield  {journal} {\bibinfo  {journal} {Nat. Commun.}\ }\textbf {\bibinfo
			{volume} {13}},\ \bibinfo {pages} {542} (\bibinfo {year} {2022})}\BibitemShut
	{NoStop}%
	\bibitem [{\citenamefont {Kim}\ \emph {et~al.}(2024)\citenamefont {Kim},
		\citenamefont {Lee}, \citenamefont {Eom}, \citenamefont {Ji}, \citenamefont
		{Choi}, \citenamefont {Joo}, \citenamefont {Bae}, \citenamefont {Kim},
		\citenamefont {Park},\ and\ \citenamefont {Park}}]{Kim2024}%
	\BibitemOpen
	\bibfield  {author} {\bibinfo {author} {\bibfnamefont {S.}~\bibnamefont
			{Kim}}, \bibinfo {author} {\bibfnamefont {H.}~\bibnamefont {Lee}}, \bibinfo
		{author} {\bibfnamefont {S.}~\bibnamefont {Eom}}, \bibinfo {author}
		{\bibfnamefont {G.}~\bibnamefont {Ji}}, \bibinfo {author} {\bibfnamefont
			{S.~H.}\ \bibnamefont {Choi}}, \bibinfo {author} {\bibfnamefont
			{H.}~\bibnamefont {Joo}}, \bibinfo {author} {\bibfnamefont {J.}~\bibnamefont
			{Bae}}, \bibinfo {author} {\bibfnamefont {K.~K.}\ \bibnamefont {Kim}},
		\bibinfo {author} {\bibfnamefont {H.-R.}\ \bibnamefont {Park}},\ and\
		\bibinfo {author} {\bibfnamefont {K.-D.}\ \bibnamefont {Park}},\ }\bibfield
	{title} {\bibinfo {title} {Dynamical control of nanoscale electron density in
			atomically thin n-type semiconductors via nano-electric pulse generator},\
	}\href {https://doi.org/10.1126/sciadv.adr0492} {\bibfield  {journal}
		{\bibinfo  {journal} {Sci. Adv.}\ }\textbf {\bibinfo {volume} {10}},\
		\bibinfo {pages} {eadr0492} (\bibinfo {year} {2024})}\BibitemShut {NoStop}%
	\bibitem [{\citenamefont {Krane}\ \emph {et~al.}(2016)\citenamefont {Krane},
		\citenamefont {Lotze}, \citenamefont {L\"{a}ger}, \citenamefont {Reecht},\
		and\ \citenamefont {Franke}}]{Krane2016}%
	\BibitemOpen
	\bibfield  {author} {\bibinfo {author} {\bibfnamefont {N.}~\bibnamefont
			{Krane}}, \bibinfo {author} {\bibfnamefont {C.}~\bibnamefont {Lotze}},
		\bibinfo {author} {\bibfnamefont {J.~M.}\ \bibnamefont {L\"{a}ger}}, \bibinfo
		{author} {\bibfnamefont {G.}~\bibnamefont {Reecht}},\ and\ \bibinfo {author}
		{\bibfnamefont {K.~J.}\ \bibnamefont {Franke}},\ }\bibfield  {title}
	{\bibinfo {title} {Electronic {Structure} and {Luminescence} of
			{Quasi}-{Freestanding} \ce{MoS_2} {Nanopatches} on {Au}(111)},\ }\href
	{https://doi.org/10.1021/acs.nanolett.6b02101} {\bibfield  {journal}
		{\bibinfo  {journal} {Nano Lett.}\ }\textbf {\bibinfo {volume} {16}},\
		\bibinfo {pages} {5163} (\bibinfo {year} {2016})}\BibitemShut {NoStop}%
	\bibitem [{\citenamefont {Pommier}\ \emph {et~al.}(2019)\citenamefont
		{Pommier}, \citenamefont {Bretel}, \citenamefont {Parra~L\'opez},
		\citenamefont {Fabre}, \citenamefont {Mayne}, \citenamefont {Boer-Duchemin},
		\citenamefont {Dujardin}, \citenamefont {Schull}, \citenamefont {Berciaud},\
		and\ \citenamefont {Le~Moal}}]{Pommier2019}%
	\BibitemOpen
	\bibfield  {author} {\bibinfo {author} {\bibfnamefont {D.}~\bibnamefont
			{Pommier}}, \bibinfo {author} {\bibfnamefont {R.}~\bibnamefont {Bretel}},
		\bibinfo {author} {\bibfnamefont {L.~E.}\ \bibnamefont {Parra~L\'opez}},
		\bibinfo {author} {\bibfnamefont {F.}~\bibnamefont {Fabre}}, \bibinfo
		{author} {\bibfnamefont {A.}~\bibnamefont {Mayne}}, \bibinfo {author}
		{\bibfnamefont {E.}~\bibnamefont {Boer-Duchemin}}, \bibinfo {author}
		{\bibfnamefont {G.}~\bibnamefont {Dujardin}}, \bibinfo {author}
		{\bibfnamefont {G.}~\bibnamefont {Schull}}, \bibinfo {author} {\bibfnamefont
			{S.}~\bibnamefont {Berciaud}},\ and\ \bibinfo {author} {\bibfnamefont
			{E.}~\bibnamefont {Le~Moal}},\ }\bibfield  {title} {\bibinfo {title}
		{Scanning tunneling microscope-induced excitonic luminescence of a
			two-dimensional semiconductor},\ }\href
	{https://doi.org/10.1103/PhysRevLett.123.027402} {\bibfield  {journal}
		{\bibinfo  {journal} {Phys. Rev. Lett.}\ }\textbf {\bibinfo {volume} {123}},\
		\bibinfo {pages} {027402} (\bibinfo {year} {2019})}\BibitemShut {NoStop}%
	\bibitem [{\citenamefont {Schuler}\ \emph {et~al.}(2020)\citenamefont
		{Schuler}, \citenamefont {Cochrane}, \citenamefont {Kastl}, \citenamefont
		{Barnard}, \citenamefont {Wong}, \citenamefont {Borys}, \citenamefont
		{Schwartzberg}, \citenamefont {Ogletree}, \citenamefont {de~Abajo},\ and\
		\citenamefont {Weber-Bargioni}}]{Schuler2020}%
	\BibitemOpen
	\bibfield  {author} {\bibinfo {author} {\bibfnamefont {B.}~\bibnamefont
			{Schuler}}, \bibinfo {author} {\bibfnamefont {K.~A.}\ \bibnamefont
			{Cochrane}}, \bibinfo {author} {\bibfnamefont {C.}~\bibnamefont {Kastl}},
		\bibinfo {author} {\bibfnamefont {E.~S.}\ \bibnamefont {Barnard}}, \bibinfo
		{author} {\bibfnamefont {E.}~\bibnamefont {Wong}}, \bibinfo {author}
		{\bibfnamefont {N.~J.}\ \bibnamefont {Borys}}, \bibinfo {author}
		{\bibfnamefont {A.~M.}\ \bibnamefont {Schwartzberg}}, \bibinfo {author}
		{\bibfnamefont {D.~F.}\ \bibnamefont {Ogletree}}, \bibinfo {author}
		{\bibfnamefont {F.~J.~G.}\ \bibnamefont {de~Abajo}},\ and\ \bibinfo {author}
		{\bibfnamefont {A.}~\bibnamefont {Weber-Bargioni}},\ }\bibfield  {title}
	{\bibinfo {title} {Electrically driven photon emission from individual atomic
			defects in monolayer \ce{WS_2}},\ }\href@noop {} {\bibfield  {journal}
		{\bibinfo  {journal} {Sci. Adv.}\ }\textbf {\bibinfo {volume} {6}} (\bibinfo
		{year} {2020})}\BibitemShut {NoStop}%
	\bibitem [{\citenamefont {P\'{e}chou}\ \emph {et~al.}(2020)\citenamefont
		{P\'{e}chou}, \citenamefont {Jia}, \citenamefont {Rigor}, \citenamefont
		{Guillermet}, \citenamefont {Seine}, \citenamefont {Lou}, \citenamefont
		{Large}, \citenamefont {Mlayah},\ and\ \citenamefont
		{Coratger}}]{Pechou2020}%
	\BibitemOpen
	\bibfield  {author} {\bibinfo {author} {\bibfnamefont {R.}~\bibnamefont
			{P\'{e}chou}}, \bibinfo {author} {\bibfnamefont {S.}~\bibnamefont {Jia}},
		\bibinfo {author} {\bibfnamefont {J.}~\bibnamefont {Rigor}}, \bibinfo
		{author} {\bibfnamefont {O.}~\bibnamefont {Guillermet}}, \bibinfo {author}
		{\bibfnamefont {G.}~\bibnamefont {Seine}}, \bibinfo {author} {\bibfnamefont
			{J.}~\bibnamefont {Lou}}, \bibinfo {author} {\bibfnamefont {N.}~\bibnamefont
			{Large}}, \bibinfo {author} {\bibfnamefont {A.}~\bibnamefont {Mlayah}},\ and\
		\bibinfo {author} {\bibfnamefont {R.}~\bibnamefont {Coratger}},\ }\bibfield
	{title} {\bibinfo {title} {Plasmonic-induced luminescence of \ce{MoSe_2}
			monolayers in a scanning tunneling microscope},\ }\href@noop {} {\bibfield
		{journal} {\bibinfo  {journal} {ACS Photonics}\ }\textbf {\bibinfo {volume}
			{7}},\ \bibinfo {pages} {3061} (\bibinfo {year} {2020})}\BibitemShut
	{NoStop}%
	\bibitem [{\citenamefont {Pe{\~n}a~Rom\'{a}n}\ \emph
		{et~al.}(2020)\citenamefont {Pe{\~n}a~Rom\'{a}n}, \citenamefont {Auad},
		\citenamefont {Grasso}, \citenamefont {Alvarez}, \citenamefont {Barcelos},\
		and\ \citenamefont {Zagonel}}]{PenaRoman2020}%
	\BibitemOpen
	\bibfield  {author} {\bibinfo {author} {\bibfnamefont {R.~J.}\ \bibnamefont
			{Pe{\~n}a~Rom\'{a}n}}, \bibinfo {author} {\bibfnamefont {Y.}~\bibnamefont
			{Auad}}, \bibinfo {author} {\bibfnamefont {L.}~\bibnamefont {Grasso}},
		\bibinfo {author} {\bibfnamefont {F.}~\bibnamefont {Alvarez}}, \bibinfo
		{author} {\bibfnamefont {I.~D.}\ \bibnamefont {Barcelos}},\ and\ \bibinfo
		{author} {\bibfnamefont {L.~F.}\ \bibnamefont {Zagonel}},\ }\bibfield
	{title} {\bibinfo {title} {Tunneling-current-induced local excitonic
			luminescence in p-doped \ce{WSe_2} monolayers},\ }\href
	{https://doi.org/10.1039/D0NR03400B} {\bibfield  {journal} {\bibinfo
			{journal} {Nanoscale}\ }\textbf {\bibinfo {volume} {12}},\ \bibinfo {pages}
		{13460} (\bibinfo {year} {2020})}\BibitemShut {NoStop}%
	\bibitem [{\citenamefont {Pe{\~n}a~Rom\'{a}n}\ \emph
		{et~al.}(2022{\natexlab{a}})\citenamefont {Pe{\~n}a~Rom\'{a}n}, \citenamefont
		{Pommier}, \citenamefont {Bretel}, \citenamefont {Parra~L\'opez},
		\citenamefont {Lorchat}, \citenamefont {Chaste}, \citenamefont {Ouerghi},
		\citenamefont {Le~Moal}, \citenamefont {Boer-Duchemin}, \citenamefont
		{Dujardin}, \citenamefont {Borisov}, \citenamefont {Zagonel}, \citenamefont
		{Schull}, \citenamefont {Berciaud},\ and\ \citenamefont
		{Le~Moal}}]{PenaRoman2022a}%
	\BibitemOpen
	\bibfield  {author} {\bibinfo {author} {\bibfnamefont {R.~J.}\ \bibnamefont
			{Pe{\~n}a~Rom\'{a}n}}, \bibinfo {author} {\bibfnamefont {D.}~\bibnamefont
			{Pommier}}, \bibinfo {author} {\bibfnamefont {R.}~\bibnamefont {Bretel}},
		\bibinfo {author} {\bibfnamefont {L.~E.}\ \bibnamefont {Parra~L\'opez}},
		\bibinfo {author} {\bibfnamefont {E.}~\bibnamefont {Lorchat}}, \bibinfo
		{author} {\bibfnamefont {J.}~\bibnamefont {Chaste}}, \bibinfo {author}
		{\bibfnamefont {A.}~\bibnamefont {Ouerghi}}, \bibinfo {author} {\bibfnamefont
			{S.}~\bibnamefont {Le~Moal}}, \bibinfo {author} {\bibfnamefont
			{E.}~\bibnamefont {Boer-Duchemin}}, \bibinfo {author} {\bibfnamefont
			{G.}~\bibnamefont {Dujardin}}, \bibinfo {author} {\bibfnamefont {A.~G.}\
			\bibnamefont {Borisov}}, \bibinfo {author} {\bibfnamefont {L.~F.}\
			\bibnamefont {Zagonel}}, \bibinfo {author} {\bibfnamefont {G.}~\bibnamefont
			{Schull}}, \bibinfo {author} {\bibfnamefont {S.}~\bibnamefont {Berciaud}},\
		and\ \bibinfo {author} {\bibfnamefont {E.}~\bibnamefont {Le~Moal}},\
	}\bibfield  {title} {\bibinfo {title} {Electroluminescence of monolayer
			\ce{WS_2} in a scanning tunneling microscope: {Effect} of bias polarity on
			spectral and angular distribution of emitted light},\ }\href
	{https://doi.org/10.1103/PhysRevB.106.085419} {\bibfield  {journal} {\bibinfo
			{journal} {Phys. Rev. B}\ }\textbf {\bibinfo {volume} {106}},\ \bibinfo
		{pages} {085419} (\bibinfo {year} {2022}{\natexlab{a}})}\BibitemShut
	{NoStop}%
	\bibitem [{\citenamefont {Ma}\ \emph {et~al.}(2022)\citenamefont {Ma},
		\citenamefont {Kalt},\ and\ \citenamefont {Stemmer}}]{Ma2022}%
	\BibitemOpen
	\bibfield  {author} {\bibinfo {author} {\bibfnamefont {Y.}~\bibnamefont
			{Ma}}, \bibinfo {author} {\bibfnamefont {R.~A.}\ \bibnamefont {Kalt}},\ and\
		\bibinfo {author} {\bibfnamefont {A.}~\bibnamefont {Stemmer}},\ }\bibfield
	{title} {\bibinfo {title} {Local strain and tunneling current modulate
			excitonic luminescence in \ce{MoS_2} monolayers},\ }\href
	{https://doi.org/10.1039/D2RA05123K} {\bibfield  {journal} {\bibinfo
			{journal} {RSC Adv.}\ }\textbf {\bibinfo {volume} {12}},\ \bibinfo {pages}
		{24922} (\bibinfo {year} {2022})}\BibitemShut {NoStop}%
	\bibitem [{\citenamefont {Parra~L\'{o}pez}\ \emph {et~al.}(2023)\citenamefont
		{Parra~L\'{o}pez}, \citenamefont {Ros{\l}awska}, \citenamefont {Scheurer},
		\citenamefont {Berciaud},\ and\ \citenamefont {Schull}}]{ParraLopez2023}%
	\BibitemOpen
	\bibfield  {author} {\bibinfo {author} {\bibfnamefont {L.~E.}\ \bibnamefont
			{Parra~L\'{o}pez}}, \bibinfo {author} {\bibfnamefont {A.}~\bibnamefont
			{Ros{\l}awska}}, \bibinfo {author} {\bibfnamefont {F.}~\bibnamefont
			{Scheurer}}, \bibinfo {author} {\bibfnamefont {S.}~\bibnamefont {Berciaud}},\
		and\ \bibinfo {author} {\bibfnamefont {G.}~\bibnamefont {Schull}},\
	}\bibfield  {title} {\bibinfo {title} {Tip-induced excitonic luminescence
			nanoscopy of an atomically resolved van der {Waals} heterostructure},\ }\href
	{https://doi.org/10.1038/s41563-023-01494-4} {\bibfield  {journal} {\bibinfo
			{journal} {Nat. Mater.}\ }\textbf {\bibinfo {volume} {22}},\ \bibinfo {pages}
		{482} (\bibinfo {year} {2023})}\BibitemShut {NoStop}%
	\bibitem [{\citenamefont {Geng}\ \emph {et~al.}(2024)\citenamefont {Geng},
		\citenamefont {Tang}, \citenamefont {Wu}, \citenamefont {Yu}, \citenamefont
		{Guest},\ and\ \citenamefont {Zhang}}]{Geng2024}%
	\BibitemOpen
	\bibfield  {author} {\bibinfo {author} {\bibfnamefont {H.}~\bibnamefont
			{Geng}}, \bibinfo {author} {\bibfnamefont {J.}~\bibnamefont {Tang}}, \bibinfo
		{author} {\bibfnamefont {Y.}~\bibnamefont {Wu}}, \bibinfo {author}
		{\bibfnamefont {Y.}~\bibnamefont {Yu}}, \bibinfo {author} {\bibfnamefont
			{J.~R.}\ \bibnamefont {Guest}},\ and\ \bibinfo {author} {\bibfnamefont
			{R.}~\bibnamefont {Zhang}},\ }\bibfield  {title} {\bibinfo {title} {Imaging
			valley excitons in a {{2D}} semiconductor with scanning tunneling
			microscope-induced luminescence},\ }\href
	{https://doi.org/10.1021/acsnano.3c12555} {\bibfield  {journal} {\bibinfo
			{journal} {ACS Nano}\ }\textbf {\bibinfo {volume} {18}},\ \bibinfo {pages}
		{8961} (\bibinfo {year} {2024})}\BibitemShut {NoStop}%
	\bibitem [{\citenamefont {Huberich}\ \emph {et~al.}(2025)\citenamefont
		{Huberich}, \citenamefont {Ammerman}, \citenamefont {Yu}, \citenamefont
		{Ren}, \citenamefont {Papadopoulos}, \citenamefont {Dong}, \citenamefont
		{Robinson}, \citenamefont {Watanabe}, \citenamefont {Taniguchi},
		\citenamefont {Gr\"{o}ning}, \citenamefont {Novotny}, \citenamefont {Li},
		\citenamefont {Wang},\ and\ \citenamefont {Schuler}}]{Huberich2025}%
	\BibitemOpen
	\bibfield  {author} {\bibinfo {author} {\bibfnamefont {L.}~\bibnamefont
			{Huberich}}, \bibinfo {author} {\bibfnamefont {E.}~\bibnamefont {Ammerman}},
		\bibinfo {author} {\bibfnamefont {G.}~\bibnamefont {Yu}}, \bibinfo {author}
		{\bibfnamefont {Y.}~\bibnamefont {Ren}}, \bibinfo {author} {\bibfnamefont
			{S.}~\bibnamefont {Papadopoulos}}, \bibinfo {author} {\bibfnamefont
			{C.}~\bibnamefont {Dong}}, \bibinfo {author} {\bibfnamefont {J.~A.}\
			\bibnamefont {Robinson}}, \bibinfo {author} {\bibfnamefont {K.}~\bibnamefont
			{Watanabe}}, \bibinfo {author} {\bibfnamefont {T.}~\bibnamefont {Taniguchi}},
		\bibinfo {author} {\bibfnamefont {O.}~\bibnamefont {Gr\"{o}ning}}, \bibinfo
		{author} {\bibfnamefont {L.}~\bibnamefont {Novotny}}, \bibinfo {author}
		{\bibfnamefont {T.}~\bibnamefont {Li}}, \bibinfo {author} {\bibfnamefont
			{S.}~\bibnamefont {Wang}},\ and\ \bibinfo {author} {\bibfnamefont
			{B.}~\bibnamefont {Schuler}},\ }\href {https://arxiv.org/abs/2510.15676}
	{\bibinfo {title} {Atomically-resolved exciton emission from single defects
			in \ce{MoS_2}}} (\bibinfo {year} {2025}),\ \Eprint
	{https://arxiv.org/abs/2510.15676} {arXiv:2510.15676 [cond-mat.mtrl-sci]}
	\BibitemShut {NoStop}%
	\bibitem [{\citenamefont {Pe{\~n}a~Rom\'{a}n}\ \emph
		{et~al.}(2022{\natexlab{b}})\citenamefont {Pe{\~n}a~Rom\'{a}n}, \citenamefont
		{Bretel}, \citenamefont {Pommier}, \citenamefont {Parra~L\'{o}pez},
		\citenamefont {Lorchat}, \citenamefont {Boer-Duchemin}, \citenamefont
		{Dujardin}, \citenamefont {Borisov}, \citenamefont {Zagonel}, \citenamefont
		{Schull}, \citenamefont {Berciaud},\ and\ \citenamefont
		{Le~Moal}}]{PenaRoman2022}%
	\BibitemOpen
	\bibfield  {author} {\bibinfo {author} {\bibfnamefont {R.~J.}\ \bibnamefont
			{Pe{\~n}a~Rom\'{a}n}}, \bibinfo {author} {\bibfnamefont {R.}~\bibnamefont
			{Bretel}}, \bibinfo {author} {\bibfnamefont {D.}~\bibnamefont {Pommier}},
		\bibinfo {author} {\bibfnamefont {L.~E.}\ \bibnamefont {Parra~L\'{o}pez}},
		\bibinfo {author} {\bibfnamefont {E.}~\bibnamefont {Lorchat}}, \bibinfo
		{author} {\bibfnamefont {E.}~\bibnamefont {Boer-Duchemin}}, \bibinfo {author}
		{\bibfnamefont {G.}~\bibnamefont {Dujardin}}, \bibinfo {author}
		{\bibfnamefont {A.~G.}\ \bibnamefont {Borisov}}, \bibinfo {author}
		{\bibfnamefont {L.~F.}\ \bibnamefont {Zagonel}}, \bibinfo {author}
		{\bibfnamefont {G.}~\bibnamefont {Schull}}, \bibinfo {author} {\bibfnamefont
			{S.}~\bibnamefont {Berciaud}},\ and\ \bibinfo {author} {\bibfnamefont
			{E.}~\bibnamefont {Le~Moal}},\ }\bibfield  {title} {\bibinfo {title}
		{Tip-induced and electrical control of the photoluminescence yield of
			monolayer \ce{WS2}},\ }\href {https://doi.org/10.1021/acs.nanolett.2c02142}
	{\bibfield  {journal} {\bibinfo  {journal} {Nano Lett.}\ }\textbf {\bibinfo
			{volume} {22}},\ \bibinfo {pages} {9244} (\bibinfo {year}
		{2022}{\natexlab{b}})}\BibitemShut {NoStop}%
	\bibitem [{\citenamefont {van Kempen}\ \emph {et~al.}(1997)\citenamefont {van
			Kempen}, \citenamefont {van Vliet}, \citenamefont {Verveer},\ and\
		\citenamefont {van~der Voort}}]{Kempen1997}%
	\BibitemOpen
	\bibfield  {author} {\bibinfo {author} {\bibfnamefont {G.~M.~P.}\
			\bibnamefont {van Kempen}}, \bibinfo {author} {\bibfnamefont {L.~J.}\
			\bibnamefont {van Vliet}}, \bibinfo {author} {\bibfnamefont {P.~J.}\
			\bibnamefont {Verveer}},\ and\ \bibinfo {author} {\bibfnamefont {H.~T.~M.}\
			\bibnamefont {van~der Voort}},\ }\bibfield  {title} {\bibinfo {title} {A
			quantitative comparison of image restoration methods for confocal
			microscopy},\ }\href {https://doi.org/10.1046/j.1365-2818.1997.d01-629.x}
	{\bibfield  {journal} {\bibinfo  {journal} {J. Microsc.}\ }\textbf {\bibinfo
			{volume} {185}},\ \bibinfo {pages} {354} (\bibinfo {year}
		{1997})}\BibitemShut {NoStop}%
	\bibitem [{\citenamefont {Verveer}\ \emph {et~al.}(1999)\citenamefont
		{Verveer}, \citenamefont {Gemkow},\ and\ \citenamefont
		{Jovin}}]{Verveer1999}%
	\BibitemOpen
	\bibfield  {author} {\bibinfo {author} {\bibfnamefont {P.~J.}\ \bibnamefont
			{Verveer}}, \bibinfo {author} {\bibfnamefont {M.~J.}\ \bibnamefont
			{Gemkow}},\ and\ \bibinfo {author} {\bibfnamefont {T.~M.}\ \bibnamefont
			{Jovin}},\ }\bibfield  {title} {\bibinfo {title} {A comparison of image
			restoration approaches applied to three-dimensional confocal and wide-field
			fluorescence microscopy},\ }\href
	{https://doi.org/10.1046/j.1365-2818.1999.00421.x} {\bibfield  {journal}
		{\bibinfo  {journal} {J. Microsc.}\ }\textbf {\bibinfo {volume} {193}},\
		\bibinfo {pages} {50} (\bibinfo {year} {1999})}\BibitemShut {NoStop}%
	\bibitem [{\citenamefont {Sibarita}(2005)}]{Sibarita2005}%
	\BibitemOpen
	\bibfield  {author} {\bibinfo {author} {\bibfnamefont {J.-B.}\ \bibnamefont
			{Sibarita}},\ }\bibinfo {title} {Deconvolution microscopy},\ in\ \href
	{https://doi.org/10.1007/b102215} {\emph {\bibinfo {booktitle} {Microscopy
				Techniques}}},\ \bibinfo {series} {Advances in Biochemical Engineering},
	Vol.~\bibinfo {volume} {95},\ \bibinfo {editor} {edited by\ \bibinfo {editor}
		{\bibfnamefont {J.}~\bibnamefont {Rietdorf}}}\ (\bibinfo  {publisher}
	{Springer Berlin Heidelberg},\ \bibinfo {address} {Berlin, Heidelberg},\
	\bibinfo {year} {2005})\ pp.\ \bibinfo {pages} {201--243}\BibitemShut
	{NoStop}%
	\bibitem [{\citenamefont {Sage}\ \emph {et~al.}(2017)\citenamefont {Sage},
		\citenamefont {Donati}, \citenamefont {Soulez}, \citenamefont {Fortun},
		\citenamefont {Schmit}, \citenamefont {Seitz}, \citenamefont {Guiet},
		\citenamefont {Vonesch},\ and\ \citenamefont {Unser}}]{Sage2017}%
	\BibitemOpen
	\bibfield  {author} {\bibinfo {author} {\bibfnamefont {D.}~\bibnamefont
			{Sage}}, \bibinfo {author} {\bibfnamefont {L.}~\bibnamefont {Donati}},
		\bibinfo {author} {\bibfnamefont {F.}~\bibnamefont {Soulez}}, \bibinfo
		{author} {\bibfnamefont {D.}~\bibnamefont {Fortun}}, \bibinfo {author}
		{\bibfnamefont {G.}~\bibnamefont {Schmit}}, \bibinfo {author} {\bibfnamefont
			{A.}~\bibnamefont {Seitz}}, \bibinfo {author} {\bibfnamefont
			{R.}~\bibnamefont {Guiet}}, \bibinfo {author} {\bibfnamefont
			{C.}~\bibnamefont {Vonesch}},\ and\ \bibinfo {author} {\bibfnamefont
			{M.}~\bibnamefont {Unser}},\ }\bibfield  {title} {\bibinfo {title}
		{Deconvolutionlab2: An open-source software for deconvolution microscopy},\
	}\href {https://doi.org/10.1016/j.ymeth.2016.12.015} {\bibfield  {journal}
		{\bibinfo  {journal} {Methods}\ }\textbf {\bibinfo {volume} {115}},\ \bibinfo
		{pages} {28} (\bibinfo {year} {2017})}\BibitemShut {NoStop}%
	\bibitem [{\citenamefont {Richardson}(1972)}]{Richardson1972}%
	\BibitemOpen
	\bibfield  {author} {\bibinfo {author} {\bibfnamefont {W.~H.}\ \bibnamefont
			{Richardson}},\ }\bibfield  {title} {\bibinfo {title} {Bayesian-based
			iterative method of image restoration},\ }\href
	{https://doi.org/10.1364/JOSA.62.000055} {\bibfield  {journal} {\bibinfo
			{journal} {J. Opt. Soc. Am.}\ }\textbf {\bibinfo {volume} {62}},\ \bibinfo
		{pages} {55} (\bibinfo {year} {1972})}\BibitemShut {NoStop}%
	\bibitem [{\citenamefont {Lucy}(1974)}]{Lucy1974}%
	\BibitemOpen
	\bibfield  {author} {\bibinfo {author} {\bibfnamefont {L.~B.}\ \bibnamefont
			{Lucy}},\ }\bibfield  {title} {\bibinfo {title} {An iterative technique for
			the rectification of observed distributions},\ }\href@noop {} {\bibfield
		{journal} {\bibinfo  {journal} {Astrophys. J.}\ }\textbf {\bibinfo {volume}
			{79}},\ \bibinfo {pages} {745} (\bibinfo {year} {1974})}\BibitemShut
	{NoStop}%
	\bibitem [{\citenamefont {Cao}\ \emph {et~al.}(2017)\citenamefont {Cao},
		\citenamefont {Le~Moal}, \citenamefont {Bigourdan}, \citenamefont {Hugonin},
		\citenamefont {Greffet}, \citenamefont {Drezet}, \citenamefont {Huant},
		\citenamefont {Dujardin},\ and\ \citenamefont {Boer-Duchemin}}]{Cao2017}%
	\BibitemOpen
	\bibfield  {author} {\bibinfo {author} {\bibfnamefont {S.}~\bibnamefont
			{Cao}}, \bibinfo {author} {\bibfnamefont {E.}~\bibnamefont {Le~Moal}},
		\bibinfo {author} {\bibfnamefont {F.}~\bibnamefont {Bigourdan}}, \bibinfo
		{author} {\bibfnamefont {J.-P.}\ \bibnamefont {Hugonin}}, \bibinfo {author}
		{\bibfnamefont {J.-J.}\ \bibnamefont {Greffet}}, \bibinfo {author}
		{\bibfnamefont {A.}~\bibnamefont {Drezet}}, \bibinfo {author} {\bibfnamefont
			{S.}~\bibnamefont {Huant}}, \bibinfo {author} {\bibfnamefont
			{G.}~\bibnamefont {Dujardin}},\ and\ \bibinfo {author} {\bibfnamefont
			{E.}~\bibnamefont {Boer-Duchemin}},\ }\bibfield  {title} {\bibinfo {title}
		{Revealing the spectral response of a plasmonic lens using low-energy
			electrons},\ }\href {https://doi.org/10.1103/PhysRevB.96.115419} {\bibfield
		{journal} {\bibinfo  {journal} {Phys. Rev. B}\ }\textbf {\bibinfo {volume}
			{96}},\ \bibinfo {pages} {115419} (\bibinfo {year} {2017})}\BibitemShut
	{NoStop}%
	\bibitem [{\citenamefont {Le~Moal}\ \emph {et~al.}(2013)\citenamefont
		{Le~Moal}, \citenamefont {Marguet}, \citenamefont {Rogez}, \citenamefont
		{Mukherjee}, \citenamefont {Dos~Santos}, \citenamefont {Boer-Duchemin},
		\citenamefont {Comtet},\ and\ \citenamefont {Dujardin}}]{LeMoal2013}%
	\BibitemOpen
	\bibfield  {author} {\bibinfo {author} {\bibfnamefont {E.}~\bibnamefont
			{Le~Moal}}, \bibinfo {author} {\bibfnamefont {S.}~\bibnamefont {Marguet}},
		\bibinfo {author} {\bibfnamefont {B.}~\bibnamefont {Rogez}}, \bibinfo
		{author} {\bibfnamefont {S.}~\bibnamefont {Mukherjee}}, \bibinfo {author}
		{\bibfnamefont {P.}~\bibnamefont {Dos~Santos}}, \bibinfo {author}
		{\bibfnamefont {E.}~\bibnamefont {Boer-Duchemin}}, \bibinfo {author}
		{\bibfnamefont {G.}~\bibnamefont {Comtet}},\ and\ \bibinfo {author}
		{\bibfnamefont {G.}~\bibnamefont {Dujardin}},\ }\bibfield  {title} {\bibinfo
		{title} {An electrically excited nanoscale light source with active angular
			control of the emitted light},\ }\href {https://doi.org/10.1021/nl401874m}
	{\bibfield  {journal} {\bibinfo  {journal} {Nano Lett.}\ }\textbf {\bibinfo
			{volume} {13}},\ \bibinfo {pages} {4198} (\bibinfo {year}
		{2013})}\BibitemShut {NoStop}%
	\bibitem [{\citenamefont {Xiao}\ \emph {et~al.}(2012)\citenamefont {Xiao},
		\citenamefont {Liu}, \citenamefont {Feng}, \citenamefont {Xu},\ and\
		\citenamefont {Yao}}]{Xiao2012}%
	\BibitemOpen
	\bibfield  {author} {\bibinfo {author} {\bibfnamefont {D.}~\bibnamefont
			{Xiao}}, \bibinfo {author} {\bibfnamefont {G.-B.}\ \bibnamefont {Liu}},
		\bibinfo {author} {\bibfnamefont {W.}~\bibnamefont {Feng}}, \bibinfo {author}
		{\bibfnamefont {X.}~\bibnamefont {Xu}},\ and\ \bibinfo {author}
		{\bibfnamefont {W.}~\bibnamefont {Yao}},\ }\bibfield  {title} {\bibinfo
		{title} {Coupled spin and valley physics in monolayers of
			${\mathrm{mos}}_{2}$ and other group-vi dichalcogenides},\ }\href
	{https://doi.org/10.1103/PhysRevLett.108.196802} {\bibfield  {journal}
		{\bibinfo  {journal} {Phys. Rev. Lett.}\ }\textbf {\bibinfo {volume} {108}},\
		\bibinfo {pages} {196802} (\bibinfo {year} {2012})}\BibitemShut {NoStop}%
	\bibitem [{\citenamefont {Cao}\ \emph {et~al.}(2012)\citenamefont {Cao},
		\citenamefont {Wang}, \citenamefont {Han}, \citenamefont {Ye}, \citenamefont
		{Zhu}, \citenamefont {Shi}, \citenamefont {Niu}, \citenamefont {Tan},
		\citenamefont {Wang}, \citenamefont {Liu},\ and\ \citenamefont
		{Feng}}]{Cao2012}%
	\BibitemOpen
	\bibfield  {author} {\bibinfo {author} {\bibfnamefont {T.}~\bibnamefont
			{Cao}}, \bibinfo {author} {\bibfnamefont {G.}~\bibnamefont {Wang}}, \bibinfo
		{author} {\bibfnamefont {W.}~\bibnamefont {Han}}, \bibinfo {author}
		{\bibfnamefont {H.}~\bibnamefont {Ye}}, \bibinfo {author} {\bibfnamefont
			{C.}~\bibnamefont {Zhu}}, \bibinfo {author} {\bibfnamefont {J.}~\bibnamefont
			{Shi}}, \bibinfo {author} {\bibfnamefont {Q.}~\bibnamefont {Niu}}, \bibinfo
		{author} {\bibfnamefont {P.}~\bibnamefont {Tan}}, \bibinfo {author}
		{\bibfnamefont {E.}~\bibnamefont {Wang}}, \bibinfo {author} {\bibfnamefont
			{B.}~\bibnamefont {Liu}},\ and\ \bibinfo {author} {\bibfnamefont
			{J.}~\bibnamefont {Feng}},\ }\bibfield  {title} {\bibinfo {title}
		{Valley-selective circular dichroism of monolayer molybdenum disulphide},\
	}\href {https://doi.org/10.1038/ncomms1882} {\bibfield  {journal} {\bibinfo
			{journal} {Nat. Commun.}\ }\textbf {\bibinfo {volume} {3}},\ \bibinfo {pages}
		{887} (\bibinfo {year} {2012})}\BibitemShut {NoStop}%
	\bibitem [{\citenamefont {Novotny}\ and\ \citenamefont
		{Hecht}(2006)}]{Novotny2006}%
	\BibitemOpen
	\bibfield  {author} {\bibinfo {author} {\bibfnamefont {L.}~\bibnamefont
			{Novotny}}\ and\ \bibinfo {author} {\bibfnamefont {B.}~\bibnamefont
			{Hecht}},\ }\href {https://doi.org/10.1017/CBO9780511813535} {\emph {\bibinfo
			{title} {Principles of {Nano}-{Optics}}}}\ (\bibinfo  {publisher} {Cambridge
		University Press},\ \bibinfo {year} {2006})\BibitemShut {NoStop}%
	\bibitem [{\citenamefont {Liu}\ \emph {et~al.}(2025)\citenamefont {Liu},
		\citenamefont {Panezai}, \citenamefont {Wang},\ and\ \citenamefont
		{Stallinga}}]{Liu2025}%
	\BibitemOpen
	\bibfield  {author} {\bibinfo {author} {\bibfnamefont {Y.}~\bibnamefont
			{Liu}}, \bibinfo {author} {\bibfnamefont {S.}~\bibnamefont {Panezai}},
		\bibinfo {author} {\bibfnamefont {Y.}~\bibnamefont {Wang}},\ and\ \bibinfo
		{author} {\bibfnamefont {S.}~\bibnamefont {Stallinga}},\ }\bibfield  {title}
	{\bibinfo {title} {Noise amplification and ill-convergence of
			{Richardson}-{Lucy} deconvolution},\ }\href
	{https://doi.org/10.1038/s41467-025-56241-x} {\bibfield  {journal} {\bibinfo
			{journal} {Nat. Commun.}\ }\textbf {\bibinfo {volume} {16}},\ \bibinfo
		{pages} {911} (\bibinfo {year} {2025})}\BibitemShut {NoStop}%
	\bibitem [{dec()}]{deconvolution}%
	\BibitemOpen
	\href@noop {} {\bibinfo {title} {{deconvolution.py}}},\ \bibinfo
	{howpublished} {GitHub,
		{https://github.com/scikit-image/scikit-image/blob/v0.25.2/skimage/restoration/deconvolution.py}
		(last accessed Nov. 14, 2025)}\BibitemShut {NoStop}%
	\bibitem [{\citenamefont {Stallinga}(2024)}]{Stallinga2024}%
	\BibitemOpen
	\bibfield  {author} {\bibinfo {author} {\bibfnamefont {S.}~\bibnamefont
			{Stallinga}},\ }\href@noop {} {\bibinfo {title}
		{{RLdeconv}\_{alldatasets.m}}},\ \bibinfo {howpublished} {GitHub,
		{https://gitlab.tudelft.nl/imphys/ci/rl-deconvolution-noise/-/blob/main/RLdeconv}\_{alldatasets.m}
		(last accessed November 14, 2025)} (\bibinfo {year} {2024})\BibitemShut
	{NoStop}%
	\bibitem [{\citenamefont {Le~Moal}\ \emph {et~al.}(2025)\citenamefont
		{Le~Moal}, \citenamefont {Laurent},\ and\ \citenamefont {Pe\~{n}a
			Rom\'{a}n}}]{LeMoal2025a}%
	\BibitemOpen
	\bibfield  {author} {\bibinfo {author} {\bibfnamefont {E.}~\bibnamefont
			{Le~Moal}}, \bibinfo {author} {\bibfnamefont {E.}~\bibnamefont {Laurent}},\
		and\ \bibinfo {author} {\bibfnamefont {R.~J.}\ \bibnamefont {Pe\~{n}a
				Rom\'{a}n}},\ }\bibfield  {title} {\bibinfo {title} {{Dataset supporting the
				publication: ``Sub-diffraction-resolved spatial distribution of emitting
				excitons in STM-induced luminescence of 2D semiconductors via Richardson-Lucy
				deconvolution''}},\ }\href {https://doi.org/10.57745/W0SEAA}
	{10.57745/W0SEAA} (\bibinfo {year} {2025})\BibitemShut {NoStop}%
	\bibitem [{\citenamefont {Castellanos-Gomez}\ \emph {et~al.}(2014)\citenamefont
		{Castellanos-Gomez}, \citenamefont {Buscema}, \citenamefont {Molenaar},
		\citenamefont {Singh}, \citenamefont {Janssen}, \citenamefont {Zant},\ and\
		\citenamefont {Steele}}]{Castellanos-Gomez2014}%
	\BibitemOpen
	\bibfield  {author} {\bibinfo {author} {\bibfnamefont {A.}~\bibnamefont
			{Castellanos-Gomez}}, \bibinfo {author} {\bibfnamefont {M.}~\bibnamefont
			{Buscema}}, \bibinfo {author} {\bibfnamefont {R.}~\bibnamefont {Molenaar}},
		\bibinfo {author} {\bibfnamefont {V.}~\bibnamefont {Singh}}, \bibinfo
		{author} {\bibfnamefont {L.}~\bibnamefont {Janssen}}, \bibinfo {author}
		{\bibfnamefont {H.~S. J. v.~d.}\ \bibnamefont {Zant}},\ and\ \bibinfo
		{author} {\bibfnamefont {G.~A.}\ \bibnamefont {Steele}},\ }\bibfield  {title}
	{\bibinfo {title} {Deterministic transfer of two-dimensional materials by
			all-dry viscoelastic stamping},\ }\href
	{http://stacks.iop.org/2053-1583/1/i=1/a=011002} {\bibfield  {journal}
		{\bibinfo  {journal} {{2D} Mater.}\ }\textbf {\bibinfo {volume} {1}},\
		\bibinfo {pages} {011002} (\bibinfo {year} {2014})}\BibitemShut {NoStop}%
	\bibitem [{\citenamefont {Mouri}\ \emph {et~al.}(2014)\citenamefont {Mouri},
		\citenamefont {Miyauchi}, \citenamefont {Toh}, \citenamefont {Zhao},
		\citenamefont {Eda},\ and\ \citenamefont {Matsuda}}]{Mouri2014}%
	\BibitemOpen
	\bibfield  {author} {\bibinfo {author} {\bibfnamefont {S.}~\bibnamefont
			{Mouri}}, \bibinfo {author} {\bibfnamefont {Y.}~\bibnamefont {Miyauchi}},
		\bibinfo {author} {\bibfnamefont {M.}~\bibnamefont {Toh}}, \bibinfo {author}
		{\bibfnamefont {W.}~\bibnamefont {Zhao}}, \bibinfo {author} {\bibfnamefont
			{G.}~\bibnamefont {Eda}},\ and\ \bibinfo {author} {\bibfnamefont
			{K.}~\bibnamefont {Matsuda}},\ }\bibfield  {title} {\bibinfo {title}
		{Nonlinear photoluminescence in atomically thin layered \ce{WSe_2} arising
			from diffusion-assisted exciton-exciton annihilation},\ }\href@noop {}
	{\bibfield  {journal} {\bibinfo  {journal} {Phys. Rev. B}\ }\textbf {\bibinfo
			{volume} {90}},\ \bibinfo {pages} {155449} (\bibinfo {year}
		{2014})}\BibitemShut {NoStop}%
	\bibitem [{\citenamefont {Kato}\ and\ \citenamefont {Kaneko}(2016)}]{Kato2016}%
	\BibitemOpen
	\bibfield  {author} {\bibinfo {author} {\bibfnamefont {T.}~\bibnamefont
			{Kato}}\ and\ \bibinfo {author} {\bibfnamefont {T.}~\bibnamefont {Kaneko}},\
	}\bibfield  {title} {\bibinfo {title} {Transport dynamics of neutral excitons
			and trions in monolayer \ce{WS2}},\ }\href@noop {} {\bibfield  {journal}
		{\bibinfo  {journal} {ACS Nano}\ }\textbf {\bibinfo {volume} {10}},\ \bibinfo
		{pages} {9687} (\bibinfo {year} {2016})}\BibitemShut {NoStop}%
	\bibitem [{\citenamefont {Kulig}\ \emph {et~al.}(2018)\citenamefont {Kulig},
		\citenamefont {Zipfel}, \citenamefont {Nagler}, \citenamefont {Blanter},
		\citenamefont {Sch\"uller}, \citenamefont {Korn}, \citenamefont {Paradiso},
		\citenamefont {Glazov},\ and\ \citenamefont {Chernikov}}]{Kulig2018}%
	\BibitemOpen
	\bibfield  {author} {\bibinfo {author} {\bibfnamefont {M.}~\bibnamefont
			{Kulig}}, \bibinfo {author} {\bibfnamefont {J.}~\bibnamefont {Zipfel}},
		\bibinfo {author} {\bibfnamefont {P.}~\bibnamefont {Nagler}}, \bibinfo
		{author} {\bibfnamefont {S.}~\bibnamefont {Blanter}}, \bibinfo {author}
		{\bibfnamefont {C.}~\bibnamefont {Sch\"uller}}, \bibinfo {author}
		{\bibfnamefont {T.}~\bibnamefont {Korn}}, \bibinfo {author} {\bibfnamefont
			{N.}~\bibnamefont {Paradiso}}, \bibinfo {author} {\bibfnamefont {M.~M.}\
			\bibnamefont {Glazov}},\ and\ \bibinfo {author} {\bibfnamefont
			{A.}~\bibnamefont {Chernikov}},\ }\bibfield  {title} {\bibinfo {title}
		{Exciton diffusion and halo effects in monolayer semiconductors},\ }\href
	{https://doi.org/10.1103/PhysRevLett.120.207401} {\bibfield  {journal}
		{\bibinfo  {journal} {Phys. Rev. Lett.}\ }\textbf {\bibinfo {volume} {120}},\
		\bibinfo {pages} {207401} (\bibinfo {year} {2018})}\BibitemShut {NoStop}%
	\bibitem [{\citenamefont {Kuhnke}\ \emph {et~al.}(2017)\citenamefont {Kuhnke},
		\citenamefont {Gro{\ss}e}, \citenamefont {Merino},\ and\ \citenamefont
		{Kern}}]{Kuhnke2017}%
	\BibitemOpen
	\bibfield  {author} {\bibinfo {author} {\bibfnamefont {K.}~\bibnamefont
			{Kuhnke}}, \bibinfo {author} {\bibfnamefont {C.}~\bibnamefont {Gro{\ss}e}},
		\bibinfo {author} {\bibfnamefont {P.}~\bibnamefont {Merino}},\ and\ \bibinfo
		{author} {\bibfnamefont {K.}~\bibnamefont {Kern}},\ }\bibfield  {title}
	{\bibinfo {title} {Atomic-{Scale} {Imaging} and {Spectroscopy} of
			{Electroluminescence} at {Molecular} {Interfaces}},\ }\href
	{https://doi.org/10.1021/acs.chemrev.6b00645} {\bibfield  {journal} {\bibinfo
			{journal} {Chem. Rev.}\ }\textbf {\bibinfo {volume} {117}},\ \bibinfo
		{pages} {5174} (\bibinfo {year} {2017})}\BibitemShut {NoStop}%
	\bibitem [{\citenamefont {Holmes}\ and\ \citenamefont
		{Liu}(1989)}]{Holmes1989}%
	\BibitemOpen
	\bibfield  {author} {\bibinfo {author} {\bibfnamefont {T.~J.}\ \bibnamefont
			{Holmes}}\ and\ \bibinfo {author} {\bibfnamefont {Y.-H.}\ \bibnamefont
			{Liu}},\ }\bibfield  {title} {\bibinfo {title} {Richardson-lucy/maximum
			likelihood image restoration algorithm for fluorescence microscopy: further
			testing},\ }\href {https://doi.org/10.1364/AO.28.004930} {\bibfield
		{journal} {\bibinfo  {journal} {Appl. Opt.}\ }\textbf {\bibinfo {volume}
			{28}},\ \bibinfo {pages} {4930} (\bibinfo {year} {1989})}\BibitemShut
	{NoStop}%
	\bibitem [{\citenamefont {Dey}\ \emph {et~al.}(2006)\citenamefont {Dey},
		\citenamefont {Blanc-Feraud}, \citenamefont {Zimmer}, \citenamefont {Roux},
		\citenamefont {Kam}, \citenamefont {Olivo-Marin},\ and\ \citenamefont
		{Zerubia}}]{Dey2006}%
	\BibitemOpen
	\bibfield  {author} {\bibinfo {author} {\bibfnamefont {N.}~\bibnamefont
			{Dey}}, \bibinfo {author} {\bibfnamefont {L.}~\bibnamefont {Blanc-Feraud}},
		\bibinfo {author} {\bibfnamefont {C.}~\bibnamefont {Zimmer}}, \bibinfo
		{author} {\bibfnamefont {P.}~\bibnamefont {Roux}}, \bibinfo {author}
		{\bibfnamefont {Z.}~\bibnamefont {Kam}}, \bibinfo {author} {\bibfnamefont
			{J.-C.}\ \bibnamefont {Olivo-Marin}},\ and\ \bibinfo {author} {\bibfnamefont
			{J.}~\bibnamefont {Zerubia}},\ }\bibfield  {title} {\bibinfo {title}
		{Richardson-lucy algorithm with total variation regularization for 3d
			confocal microscope deconvolution},\ }\href
	{https://doi.org/10.1002/jemt.20294} {\bibfield  {journal} {\bibinfo
			{journal} {Microsc. Res. Techniq.}\ }\textbf {\bibinfo {volume} {69}},\
		\bibinfo {pages} {260} (\bibinfo {year} {2006})}\BibitemShut {NoStop}%
	\bibitem [{\citenamefont {Laasmaa}\ \emph {et~al.}(2011)\citenamefont
		{Laasmaa}, \citenamefont {Vendelin},\ and\ \citenamefont
		{Peterson}}]{Laasmaa2011}%
	\BibitemOpen
	\bibfield  {author} {\bibinfo {author} {\bibfnamefont {M.}~\bibnamefont
			{Laasmaa}}, \bibinfo {author} {\bibfnamefont {M.}~\bibnamefont {Vendelin}},\
		and\ \bibinfo {author} {\bibfnamefont {P.}~\bibnamefont {Peterson}},\
	}\bibfield  {title} {\bibinfo {title} {Application of {Regularized}
			{Richardson}-{Lucy} {Algorithm} for {Deconvolution} of {Confocal}
			{Microscopy} {Images}},\ }\href {https://doi.org/10.1016/j.bpj.2010.12.963}
	{\bibfield  {journal} {\bibinfo  {journal} {Biophys. J.}\ }\textbf {\bibinfo
			{volume} {100}},\ \bibinfo {pages} {139a} (\bibinfo {year}
		{2011})}\BibitemShut {NoStop}%
	\bibitem [{\citenamefont {Mukamel}\ \emph {et~al.}(2012)\citenamefont
		{Mukamel}, \citenamefont {Babcock},\ and\ \citenamefont
		{Zhuang}}]{Mukamel2012}%
	\BibitemOpen
	\bibfield  {author} {\bibinfo {author} {\bibfnamefont {E.}~\bibnamefont
			{Mukamel}}, \bibinfo {author} {\bibfnamefont {H.}~\bibnamefont {Babcock}},\
		and\ \bibinfo {author} {\bibfnamefont {X.}~\bibnamefont {Zhuang}},\
	}\bibfield  {title} {\bibinfo {title} {Statistical {Deconvolution} for
			{Superresolution} {Fluorescence} {Microscopy}},\ }\href
	{https://doi.org/10.1016/j.bpj.2012.03.070} {\bibfield  {journal} {\bibinfo
			{journal} {Biophys. J.}\ }\textbf {\bibinfo {volume} {102}},\ \bibinfo
		{pages} {2391} (\bibinfo {year} {2012})}\BibitemShut {NoStop}%
	\bibitem [{\citenamefont {Wang}\ \emph {et~al.}(2014)\citenamefont {Wang},
		\citenamefont {Meza}, \citenamefont {Wang}, \citenamefont {Gao},\ and\
		\citenamefont {Liu}}]{Wang2014b}%
	\BibitemOpen
	\bibfield  {author} {\bibinfo {author} {\bibfnamefont {D.}~\bibnamefont
			{Wang}}, \bibinfo {author} {\bibfnamefont {D.}~\bibnamefont {Meza}}, \bibinfo
		{author} {\bibfnamefont {Y.}~\bibnamefont {Wang}}, \bibinfo {author}
		{\bibfnamefont {L.}~\bibnamefont {Gao}},\ and\ \bibinfo {author}
		{\bibfnamefont {J.~T.~C.}\ \bibnamefont {Liu}},\ }\bibfield  {title}
	{\bibinfo {title} {Sheet-scanned dual-axis confocal microscopy using
			richardson-lucy deconvolution},\ }\href
	{https://doi.org/10.1364/OL.39.005431} {\bibfield  {journal} {\bibinfo
			{journal} {Opt. Lett.}\ }\textbf {\bibinfo {volume} {39}},\ \bibinfo {pages}
		{5431} (\bibinfo {year} {2014})}\BibitemShut {NoStop}%
	\bibitem [{\citenamefont {Str\"{o}hl}\ and\ \citenamefont
		{Kaminski}(2015)}]{Stroehl2015}%
	\BibitemOpen
	\bibfield  {author} {\bibinfo {author} {\bibfnamefont {F.}~\bibnamefont
			{Str\"{o}hl}}\ and\ \bibinfo {author} {\bibfnamefont {C.~F.}\ \bibnamefont
			{Kaminski}},\ }\bibfield  {title} {\bibinfo {title} {A joint richardson-lucy
			deconvolution algorithm for the reconstruction of multifocal structured
			illumination microscopy data},\ }\href
	{https://doi.org/10.1088/2050-6120/3/1/014002} {\bibfield  {journal}
		{\bibinfo  {journal} {Methods Appl. Fluoresc.}\ }\textbf {\bibinfo {volume}
			{3}},\ \bibinfo {pages} {014002} (\bibinfo {year} {2015})}\BibitemShut
	{NoStop}%
	\bibitem [{\citenamefont {Perez}\ \emph {et~al.}(2016)\citenamefont {Perez},
		\citenamefont {Chang},\ and\ \citenamefont {Stelzer}}]{Perez2016}%
	\BibitemOpen
	\bibfield  {author} {\bibinfo {author} {\bibfnamefont {V.}~\bibnamefont
			{Perez}}, \bibinfo {author} {\bibfnamefont {B.-J.}\ \bibnamefont {Chang}},\
		and\ \bibinfo {author} {\bibfnamefont {E.~H.~K.}\ \bibnamefont {Stelzer}},\
	}\bibfield  {title} {\bibinfo {title} {Optimal {{2D}}-{SIM} reconstruction by
			two filtering steps with {Richardson}-{Lucy} deconvolution},\ }\href
	{https://doi.org/10.1038/srep37149} {\bibfield  {journal} {\bibinfo
			{journal} {Sci. Rep.}\ }\textbf {\bibinfo {volume} {6}},\ \bibinfo {pages}
		{37149} (\bibinfo {year} {2016})}\BibitemShut {NoStop}%
	\bibitem [{\citenamefont {Zhang}\ \emph {et~al.}(2019)\citenamefont {Zhang},
		\citenamefont {Lang}, \citenamefont {Wang}, \citenamefont {Liao},\ and\
		\citenamefont {Gong}}]{Zhang2019}%
	\BibitemOpen
	\bibfield  {author} {\bibinfo {author} {\bibfnamefont {Y.}~\bibnamefont
			{Zhang}}, \bibinfo {author} {\bibfnamefont {S.}~\bibnamefont {Lang}},
		\bibinfo {author} {\bibfnamefont {H.}~\bibnamefont {Wang}}, \bibinfo {author}
		{\bibfnamefont {J.}~\bibnamefont {Liao}},\ and\ \bibinfo {author}
		{\bibfnamefont {Y.}~\bibnamefont {Gong}},\ }\bibfield  {title} {\bibinfo
		{title} {Super-resolution algorithm based on richardson--lucy deconvolution
			for three-dimensional structured illumination microscopy},\ }\href
	{https://doi.org/10.1364/JOSAA.36.000173} {\bibfield  {journal} {\bibinfo
			{journal} {J. Opt. Soc. Am. A}\ }\textbf {\bibinfo {volume} {36}},\ \bibinfo
		{pages} {173} (\bibinfo {year} {2019})}\BibitemShut {NoStop}%
	\bibitem [{\citenamefont {Schneider}\ \emph {et~al.}(2012)\citenamefont
		{Schneider}, \citenamefont {Rasband},\ and\ \citenamefont
		{Eliceiri}}]{Schneider2012}%
	\BibitemOpen
	\bibfield  {author} {\bibinfo {author} {\bibfnamefont {C.~A.}\ \bibnamefont
			{Schneider}}, \bibinfo {author} {\bibfnamefont {W.~S.}\ \bibnamefont
			{Rasband}},\ and\ \bibinfo {author} {\bibfnamefont {K.~W.}\ \bibnamefont
			{Eliceiri}},\ }\bibfield  {title} {\bibinfo {title} {{NIH} {Image} to
			{ImageJ}: 25 years of image analysis},\ }\href
	{https://doi.org/10.1038/nmeth.2089} {\bibfield  {journal} {\bibinfo
			{journal} {Nat. Methods}\ }\textbf {\bibinfo {volume} {9}},\ \bibinfo {pages}
		{671} (\bibinfo {year} {2012})}\BibitemShut {NoStop}%
	\bibitem [{\citenamefont {Morozov}\ \emph {et~al.}(2021)\citenamefont
		{Morozov}, \citenamefont {Wolff},\ and\ \citenamefont
		{Mortensen}}]{Morozov2021}%
	\BibitemOpen
	\bibfield  {author} {\bibinfo {author} {\bibfnamefont {S.}~\bibnamefont
			{Morozov}}, \bibinfo {author} {\bibfnamefont {C.}~\bibnamefont {Wolff}},\
		and\ \bibinfo {author} {\bibfnamefont {N.~A.}\ \bibnamefont {Mortensen}},\
	}\bibfield  {title} {\bibinfo {title} {Room-temperature low-voltage control
			of excitonic emission in transition metal dichalcogenide monolayers},\ }\href
	{https://doi.org/10.1002/adom.202101305} {\bibfield  {journal} {\bibinfo
			{journal} {Adv. Opt. Mater.}\ }\textbf {\bibinfo {volume} {9}},\ \bibinfo
		{pages} {2101305} (\bibinfo {year} {2021})}\BibitemShut {NoStop}%
	\bibitem [{\citenamefont {Brito}\ \emph {et~al.}(2024)\citenamefont {Brito},
		\citenamefont {Costa}, \citenamefont {Ceccatto}, \citenamefont {de~Almeida},
		\citenamefont {de~Siervo}, \citenamefont {Couto}, \citenamefont {Barcelos},\
		and\ \citenamefont {Zagonel}}]{Brito2024}%
	\BibitemOpen
	\bibfield  {author} {\bibinfo {author} {\bibfnamefont {T.~G.~L.}\
			\bibnamefont {Brito}}, \bibinfo {author} {\bibfnamefont {F.~J.~R.}\
			\bibnamefont {Costa}}, \bibinfo {author} {\bibfnamefont {A.}~\bibnamefont
			{Ceccatto}}, \bibinfo {author} {\bibfnamefont {C.~A.~N.}\ \bibnamefont
			{de~Almeida}}, \bibinfo {author} {\bibfnamefont {A.}~\bibnamefont
			{de~Siervo}}, \bibinfo {author} {\bibfnamefont {O.~D.~D.}\ \bibnamefont
			{Couto}}, \bibinfo {author} {\bibfnamefont {I.~D.}\ \bibnamefont
			{Barcelos}},\ and\ \bibinfo {author} {\bibfnamefont {L.~F.}\ \bibnamefont
			{Zagonel}},\ }\bibfield  {title} {\bibinfo {title} {Investigating the impact
			of ito substrates on the optical and electronic properties of \ce{WSe_2}
			monolayers},\ }\href {https://doi.org/10.1088/1361-6528/ad8fb4} {\bibfield
		{journal} {\bibinfo  {journal} {Nanotechnology}\ }\textbf {\bibinfo {volume}
			{36}},\ \bibinfo {pages} {055704} (\bibinfo {year} {2024})}\BibitemShut
	{NoStop}%
	\bibitem [{\citenamefont {Lucy}(1992{\natexlab{a}})}]{Lucy1992}%
	\BibitemOpen
	\bibfield  {author} {\bibinfo {author} {\bibfnamefont {L.~B.}\ \bibnamefont
			{Lucy}},\ }\bibfield  {title} {\bibinfo {title} {Statistical limits to super
			resolution},\ }\href@noop {} {\bibfield  {journal} {\bibinfo  {journal}
			{Astron. Astrophys.}\ }\textbf {\bibinfo {volume} {261}},\ \bibinfo {pages}
		{706} (\bibinfo {year} {1992}{\natexlab{a}})}\BibitemShut {NoStop}%
	\bibitem [{\citenamefont {Lucy}(1992{\natexlab{b}})}]{Lucy1992a}%
	\BibitemOpen
	\bibfield  {author} {\bibinfo {author} {\bibfnamefont {L.~B.}\ \bibnamefont
			{Lucy}},\ }\bibfield  {title} {\bibinfo {title} {Resolution limits for
			deconvolved images},\ }\href@noop {} {\bibfield  {journal} {\bibinfo
			{journal} {Astrophys. J.}\ }\textbf {\bibinfo {volume} {104}},\ \bibinfo
		{pages} {1260} (\bibinfo {year} {1992}{\natexlab{b}})}\BibitemShut {NoStop}%
\end{thebibliography}

%

\end{document}



\title{Supplemental Material for: Sub-diffraction-resolved spatial distribution of emitting excitons in STM-induced luminescence of 2D semiconductors via Richardson-Lucy deconvolution}



\author{Elys\'{e} Laurent} 
\affiliation{Universit\'{e} Paris-Saclay, CNRS, Institut des Sciences Mol\'{e}culaires d'Orsay, 91405, Orsay, France}
\author{Ricardo Javier Pe\~{n}a Rom\'{a}n}
\affiliation{Institute of Physics ``Gleb Wataghin'', Department of Applied Physics, State University of Campinas-UNICAMP, 13083-859, Campinas, Brazil}
\altaffiliation{Max Planck Institute for Solid State Research, Stuttgart 70569, Germany}
\altaffiliation{Technical University of Dresden, Institute of Solid State and Materials Physics,  Dresden 01069, Germany}
\author{Sarah Miller}
\affiliation{Universit\'{e} Paris-Saclay, CNRS, Institut des Sciences Mol\'{e}culaires d'Orsay, 91405, Orsay, France}
\author{Aditi Raman Moghe}
\affiliation{Institut de Physique et de Chimie des Mat\'{e}riaux de Strasbourg, Universit\'{e} de Strasbourg, CNRS, IPCMS, UMR 7504, F-67000 Strasbourg, France}
\author{Etienne Lorchat}
\affiliation{NTT Research, Inc., Physics \& Informatics (PHI) Laboratories, Sunnyvale, CA 94085, USA}
\author{S\'{e}verine Le Moal}
\author{Elizabeth Boer-Duchemin} 
\affiliation{Universit\'{e} Paris-Saclay, CNRS, Institut des Sciences Mol\'{e}culaires d'Orsay, 91405, Orsay, France}
\author{Luiz Fernando Zagonel}
\affiliation{Institute of Physics ``Gleb Wataghin'', Department of Applied Physics, State University of Campinas-UNICAMP, 13083-859, Campinas, Brazil}
\author{St\'{e}phane Berciaud}
\affiliation{Institut de Physique et de Chimie des Mat\'{e}riaux de Strasbourg, Universit\'{e} de Strasbourg, CNRS, IPCMS, UMR 7504, F-67000 Strasbourg, France}
\author{Eric Le Moal}
\affiliation{Universit\'{e} Paris-Saclay, CNRS, Institut des Sciences Mol\'{e}culaires d'Orsay, 91405, Orsay, France}
\email{eric.le-moal@universite-paris-saclay.fr}


\date{November 28, 2025}

	


\maketitle



%
\begin{figure}
	\includegraphics[width=0.6\linewidth]{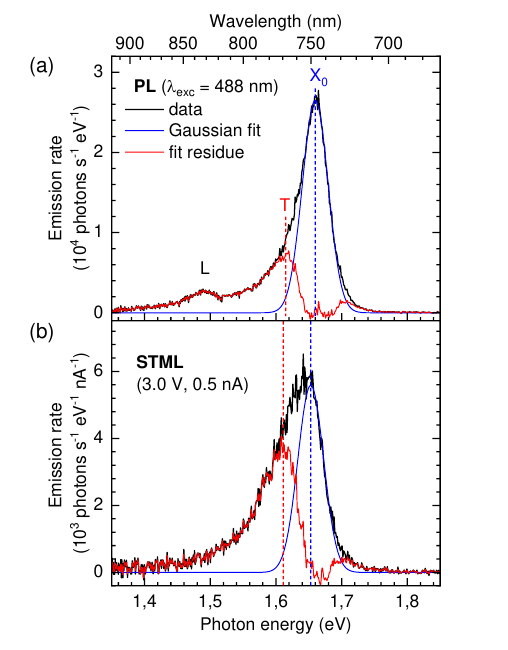} 
	\caption{Comparison of the photoluminescence (PL) and STM-induced luminescence (STML) spectra measured in the same area of the sample. (a) PL spectrum: continuous-wave laser excitation at a wavelength of $488$~nm, i.e., a photon energy of $2.54$~eV (wide-field illumination in normal incidence, laser power $0.01$~W~cm$^{-2}$, acquisition time $60$~s). (b) STML spectrum: sample bias $3.0$~V, current setpoint $0.5$~nA, acquisition time $120$~s. The same grating ($150$~lines/mm) and the same slit width ($40~\mu$m) were used for the PL and STML spectra. (a) and (b) Black curve: experimental data corrected for the detection efficiency of our setup. Blue curve: fit of the high-energy side of the emission peak with a Gaussian function; this contribution is assigned to neutral excitons ($X_0$). Red curve: residue of the fit, which highlight contributions of charged excitons (trions, $T$) and localized or defect-bound excitons ($L$). The latter contribution is not present in the STML spectrum because the tunneling-induced excitation is spatially much more localized. }
	\label{FIG-S1}
\end{figure}

%
\begin{figure}
	\includegraphics[width=1.0\linewidth]{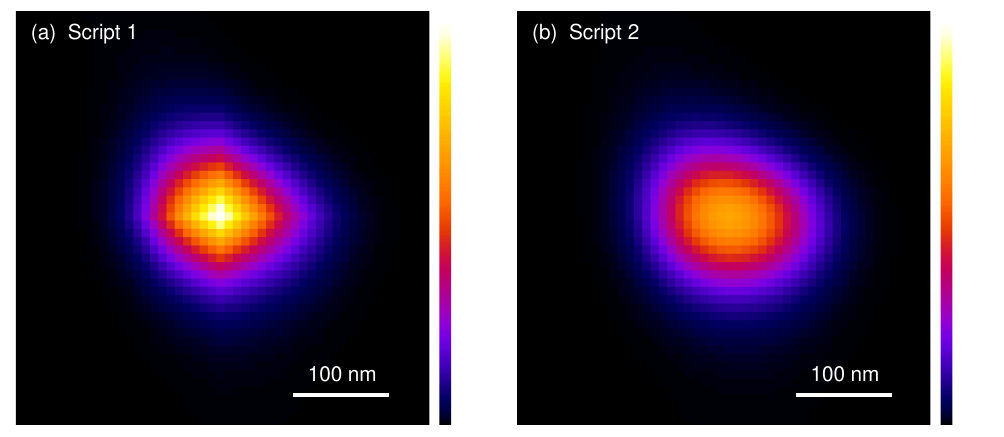} 
	\caption{Zoom of the two deconvolved images shown in Figure~1(h) of the article, which were obtained using two different versions of the Richardson-Lucy (RL) algorithm. Script~1 used in (a) requires computing 2D convolution between object and point spread function (PSF) at each iteration, while in script~2 used in (b) this operation is replaced by the product of their Fourier transform (see Methods in the article). A convergence issue is found for the RL algorithm using script~1, which yields that the central spot in the deconvolved image tends to take a cross shape beyond a certain number of iterations.}
	\label{FIG-S2}
\end{figure}